\documentclass[]{emulateapj}
\usepackage{apjfonts}
\pdfoutput=1
\usepackage{threeparttable}
\usepackage{textcomp}

\usepackage{url}
\usepackage{CJKutf8}
\shorttitle{K+A Galaxies in the Cl1604 Supercluster}

\begin{document}

\title{Star Formation Quenching in High-redshift Large-scale Structure: Post-starburst Galaxies in the Cl1604 Supercluster at $z \sim 0.9$}

\author{Po-Feng Wu \begin{CJK*}{UTF8}{bsmi}(吳柏鋒)\end{CJK*}\altaffilmark{1}, Roy R. Gal\altaffilmark{1}, Brian C. Lemaux\altaffilmark{2}, Dale D. Kocevski\altaffilmark{3}, Lori M. Lubin\altaffilmark{4}, Nicholas Rumbaugh\altaffilmark{4}, and Gordon K. Squires\altaffilmark{5}}

\altaffiltext{1}{University of Hawaii, Institute for Astronomy, 2680 Woodlawn Drive, HI 96822, USA}
\altaffiltext{2}{Aix Marseille Universit\'{e}, CNRS, LAM (Laboratoire d\textasciiacute Astrophysique de Marseille), UMR 7326, 13388, Marseille, France}
\altaffiltext{3}{University of Kentucky, Department of Physics and Astronomy, Lexington, KY 40506, USA}
\altaffiltext{4}{Department of Physics, University of California-Davis, 1 Shields Avenue, Davis, CA 95616, USA}
\altaffiltext{5}{California Institute of Technology, M/S 220-6, 1200 E. California Boulevard, Pasadena CA 91125, USA}

\begin{abstract}

The Cl1604 supercluster at $z \sim 0.9$ is one of the most extensively studied high redshift large scale structures, with more than 500 spectroscopically confirmed members. It consists of 8 clusters and groups, with members numbering from a dozen to nearly a hundred, providing a broad range of environments for investigating the large scale environmental effects on galaxy evolution. Here we examine the properties of 48 post-starburst galaxies in Cl1604, comparing them to other galaxy populations in the same supercluster.  Incorporating photometry from ground-based optical and near-infrared imaging, along with $Spitzer$ mid-infrared observations, we derive stellar masses for all Cl1604 members.  The colors and stellar masses of the K+A galaxies support the idea that they are progenitors of red sequence galaxies. Their morphologies, residual star-formation rates, and spatial distributions suggest galaxy mergers may be the principal mechanism producing post-starburst galaxies. Interaction between galaxies and 
the dense intra-cluster medium is also effective, but only in the cores of dynamically evolved clusters. 
The prevalence of post-starburst galaxies in clusters correlates with the dynamical state of the host cluster, as both galaxy mergers and the dense intra-cluster medium produce post-starburst galaxies. We also investigate the incompleteness and contamination of K+A samples selected by means of H$\delta$ and [OII] equivalent widths. K+A samples may be up to $\sim50\%$ incomplete due to the presence of LINER/Seyferts and up to $\sim30\%$ of K+A galaxies could have substantial star formation activity. 

\end{abstract}


\section{INTRODUCTION}

In recent years, large galaxy surveys have greatly enhanced our understanding of galaxy properties. One of the most striking results is the strongly bimodal distribution of galaxy colors \citep{str01,bal04}. There are two distinct populations: the so-called red-sequence, dominated by galaxies with quiescent star-formation and old stellar populations, and the ``blue cloud``, containing mainly star-forming, morphologically disky galaxies. This bimodal color distribution is found to exist not only locally, but out to at least $z \sim 1.5$ as well \citep{bel04,wil06,fra07}. This clear separation in color has thus existed for a significant fraction of the universe's age, but the number density in each population is seen to evolve differently. Since $z \sim 1$, the number density of blue galaxies has remained nearly unchanged, while the number density of red galaxies has increased by a factor of 2 to 4 \citep{fab07}. 
These simple but striking results have immediate important implications for galaxy evolution. First, at least some blue, star-forming galaxies must transform into passive red galaxies, so that the red sequence is built up over time. Second, the transition phase can last for only a short fraction of the lifespan of a galaxy, so that only a few galaxies are observed in their transition phase. In other words, the star-formation rate should decline rapidly, via some quenching process, during a galaxy's transition from the blue to the red population. 

The natural question is, what causes this quenching? Recently, using the SDSS and zCOSMOS surveys, \citet{pen10} demonstrated that star formation quenching is controlled by two separable factors: the galaxy stellar mass and the environment, at least out to $z \sim 1$, where galaxies with higher stellar masses and residing in denser regions have a higher likelihood of being quenched. A number of specific quenching mechanisms have been proposed. To create the galaxy stellar mass dependence, one mechanism is shock heating of gas falling into a massive dark matter halo. In halos with masses above $\sim 10^{12} M_{\odot}$, infalling cold gas would be shock heated to the virial temperature, preventing the formation of new stars \citep{dek03}. Additionally, active galactic nuclei (AGN) or supernovae could inject energy to heat up or blow out the gas, further suppressing star formation \citep{spi05,cro06,kav07}. 
Environmental dependent mechanisms that can quench star-formation also have been proposed. Galaxy mergers and interactions can induce a short period of elevated star formation \citep{mih94,bar92}, exhausting the available gas and resulting in a rapid decline of the star formation rate \citep{sny11}. For example, in galaxy clusters, interstellar gas in galaxies may be removed by interaction with the dense intra-cluster medium (ICM), known as ram-pressure stripping \citep{gun72,aba99,tre03}.

Direct observations of galaxies which have recently undergone quenching may reveal the quenching mechanisms at work. However, one of the difficulties in studying this quenching, or in general, galaxy evolution, is that, we only see a snapshot of a galaxy in the time domain, and obtaining the true star formation history of a galaxy is a non-trivial task. Luckily, post-starburst galaxies provide an opening for such studies, since their properties indicate a relatively specific time frame when they experienced quenching. These galaxies are selected spectroscopically, exhibiting strong Balmer absorption lines, indicating the presence of A stars, and a lack of ionized gas producing H$\alpha$ or other emission lines, implying no on-going star formation. This combination of spectral features implies an episode of intense star formation activity roughly a billion years ago which was subsequently quenched, so that the galaxy does not have OB stars which emit ionizing photons but contains a significant population of 
A stars. This type of spectral feature fades away quickly, within the lifetime of A stars, $\lesssim$ 1~Gyr. As a result, post-starburst galaxies provide a proxy to assess the physical conditions close to the time that quenching happened, and they have often been suggested to be progenitors of early-type galaxies \citep{tra04,qui04,got07a,ma08}.  Although their name implies that they experienced a starburst, or elevated SFR, before quenching, such a star formation history is not necessary to produce strong Balmer absorption lines. As shown by \citet{leb06}, any rapidly declining star formation history can produce ``post-starburst``-type spectra. Because a burst is not necessary, \citet{yan04} proposed ``post-quenching`` galaxies as a more appropriate way to describe this spectral type. These galaxies are also called E+A or K+A galaxies, where E indicates a typical early-type galaxy spectrum and K represents a stellar population dominated by K-type stars \citep{dre83,bal99,dre99}.

Finding post-starburst galaxies is not an easy task. Almost by definition, post-starburst galaxies comprise only a small fraction of the whole galaxy population. At $z \sim 0.1$, only $\sim 0.1-0.2\%$ of galaxies are in their post-starburst phase \citep{bla04,got05,got07b,yan09}. Even with such a low prevalence in the nearby universe, SDSS, with its enormous spectroscopic sample, has identified more than a thousand post-starburst galaxies \citep{qui04,cho09}. But the situation becomes more difficult at higher redshifts because obtaining spectra of galaxies at ever greater distances incurs ever more observational resources. Nevertheless, surveys at intermediate redshifts have shown that the prevalence of post-starburst galaxies are generally higher \citep[$\sim 1\%$,][]{tra04,yan09}. Even with these projects, relatively few post-starburst galaxies had been found. 

Recently, large redshift surveys such as zCOSMOS \citep{lil07} and DEEP2 \citep{dav03} have obtained thousands of galaxy spectra up to $z \sim 1$. A few dozen post-starburst galaxies were identified from the large number of spectra, and analysis of their mass and environmental dependence beyond the nearby universe becomes possible \citep{yan09,ver10}. Nevertheless, galaxies in these field surveys largely reside in typical (field) environments. High density regions only comprise a small fraction of the total survey volume, so that post-starburst galaxies from these surveys are not indicative of the environmental effects at play in the high density universe.

On the other hand, a few spectroscopic cluster surveys specifically targeting intermediate to high redshift clusters have obtained hundreds to a thousand spectra of cluster galaxies, e.g., ICBS at $ 0.31 < z < 0.54 $ \citep{oem13}, EDisCS at $ 0.4 < z < 0.8$ \citep{pog06}, and GCLASS at $0.85 < z < 1.2$ \citep{muz12}. They provide complementary data on post-starburst galaxies in dense regions of the universe. Generally, post-starburst galaxies are at least a few times more prevalent in clusters than in the field at similar redshifts. The prevalence in these intermediate to high redshift galaxy groups and clusters reaches 10\% or even higher \citep{tra03,pog09,muz12,dre13}, with a trend that denser regions and more massive clusters have higher post-starburst fractions \citep{pog09,dre13}, but large variations are seen among systems. Although the prevalence of post-starburst galaxies in high density regions shows that environmental effects play an important role, it is still unclear what environmental-
dependent 
mechanisms are responsible. Several investigations have been undertaken, but a consensus on which are dominant has not yet been achieved \citep[][to name a few]{pog99,tra04,pra05,ma08,oem09,sny11,dre13}. 

In this work, we examine high redshift large-scale structures, where quenching is actively under way. We focus on post-starburst galaxies in the Cl1604 supercluster at $z \sim 0.9$ \citep{gal08,koc09,lem10,koc11a,lem12,rum13}, the most extensively studied large-scale structure in the Observation of Redshift Evolution in Large Scale Environments (ORELSE) survey \citep{lub09}. This structure spans 13 Mpc $\times$ 100 Mpc and consists of at least 8 individual clusters and groups. This field is covered by multi-wavelength observations, including HST imaging. More than 500 galaxies have been spectroscopically confirmed in the whole structure, which is one of the largest samples in any high density region at similar redshift. For each system, the number of confirmed members ranges from a few to $\sim$100, with velocity dispersions $\sigma_v \simeq 300 \sim 800$ km s$^{-1}$ \citep{lub00,gal04,gal05,gal08,lem12}. Using post-starburst galaxies as proxies, the broad range of properties allow us to 
investigate how the large-scale structure affects galaxy evolution, and what may be the physical mechanism(s) responsible for producing post-starburst galaxies. 

In this paper, we will use the terminology of ``K+A galaxy`` for most of our discussion, as our sample is selected based solely on spectral features. In Section 2, we describe the data used in the paper and the data reduction. In Section 3, we define our K+A sample, cluster and group membership, and derive stellar masses for the galaxies. Properties of K+A galaxies are presented in Section 4. Implications from the results are discussed in Section 5. Throughout this paper we assume a $\Lambda$CDM cosmology with $\Omega_M = 0.27$, $\Omega_{\Lambda}=0.73$, and $H_0 = 70$ km s$^{-1}$ Mpc$^{-1}$. Magnitudes are given in the AB system.

\section{OBSERVATIONS}

\subsection{Optical Imaging}
Our ground-based optical photometry consists of two pointings of the Large Format Camera \citep[LFC;][]{sim00} in Sloan $r'$, $i'$, and $z'$ and two pointings of the Carnegie Observatories Spectroscopic Multislit and Imaging Camera \citep[COSMIC;][]{kel98} in Cousins $R$ and Gunn $i$ filters, both on the Palomar 5m telescope. The layout of the observations can be found in Figure~4 of \citet{gal08}. These data were reduced using the Image Reduction and Analysis Facility \citep[IRAF, ][]{tod86} with a set of publicly available routines. The detailed data reduction process is described in \citet{gal05}. The 5 $\sigma$ limiting magnitudes are 25.2, 24.8, and 23.3 in $r'$, $i'$, and $z'$, respectively. However, we have improved the source detection process from the previous work, as detailed in \S2.6. 

\subsection{Near-infrared Imaging}
\hyphenation{WFCAM}
Near-infrared (NIR) $J$ and $K_s$ band images were obtained with the Wide-Field Camera \citep[WFCAM;][]{cas07} on the United Kingdom Infrared Telescope (UKIRT). This camera has four widely-separated detectors, and four tiled pointings completely fill in a square with $0.75^{\circ}$ sides. Observations were taken between 2007 and 2010, in $\sim0.9"$ seeing. Images were taken with individual exposure times of 20s in $J$ and 10s in $K_s$, with microstepping among exposures to recover some of the resolution lost to WFCAM's large pixels. Total integration times per pixel were 1500s and 1800s in $J$ and $K_s$, respectively. Data were processed by the WFCAM reduction pipeline at the Cambridge Astronomy Survey Unit. The ``deepleavestacks`` for each pointing were retrieved from the WFCAM science archive, and combined using the software task {\it SWarp} \citep{ber02}.

\subsection{HST Imaging}
Part of the Cl1604 field is covered by a 17 pointing \textit{Hubble Space Telescope (HST)} Advanced Camera for Surveys (ACS) mosaic in the $F606W$ and $F814W$ filters. Observations consist of 2 pointings from GO-9919 (PI: H. C. Ford) with effective exposure times of 4840s and 15 pointings from GO-11003 (PI: L. M. Lubin) with effective exposure times of 1998s. The integration time of 1998s results in a 5$\sigma$ point source limiting magnitudes of 27.2 and 26.8 mag in $F606W$ and $F814W$, respectively. In terms of physical units, it corresponds to 94.0 $L_{\sun}$ pc$^{-2}$ \citep{asc13}. Detailed descriptions of the observations, data reduction and photometry are provided in \citet{koc11a}. In our study, the \textit{HST} data are used to obtain optical galaxy colors, examine blended sources in LFC images, and determine galaxy morphologies. 

\subsection{Spitzer IRAC and MIPS Data}
The entire Cl1604 field was imaged by the \textit{Spitzer} Infrared Array Camera (IRAC) at 3.6 and 4.5 $\mu$m and by the Multiband Imaging Spectrometer (MIPS) at 24$\mu$m. For the observational program, we refer the reader to \citet{koc11a}. Data were reduced using the standard \textit{Spitzer} Science Center (SSC) reduction pipeline. For IRAC images, the reduced basic calibrated data (BCD) images were further processed using a modified version of the SWIRE survey pipeline \citep{sur05a} to remove instrumental artifacts. The processed images were co-added using the SSC MOsaicker and Point source EXtractor (MOPEX)  version 18.4.9. We created the Fiducial Image Frame (FIF) using both IRAC channels, resulting in an unified astrometric reference for both channels. The permanently damaged pixels of the detector and bad pixels in each BCD were masked out when co-adding. The pixel scale was set to 0.\arcsec6 pixel$^{-1}$. 

The MIPS data reduction is described in \citet{koc11a}. With a total exposure time of 1200s pixel$^{-1}$, we achieved a 3$\sigma$ flux limit of 40 $\mu$Jy. Following the recipe and the synthetic spectra of \citet{cha01} and \citet{dal02}, we calculated the total IR luminosity ($L_{TIR}$; 8--1000 $\mu$m) of each 24$\mu$m source. The SFR can be directly obtained from $L_{TIR}$ by the $L_{TIR}$--SFR relation from \citet{ken98}. Our 40$\mu$Jy flux limit corresponds to a $L_{TIR}$ of $3 \times  10^{10} L_{\odot}$ at $z = 0.9$, an SFR of 5.2 $M_{\odot}$ yr$^{-1}$ \citep{koc11a}. 

\subsection{Optical Spectroscopy}

The spectroscopic data were obtained using the Low-Resolution Imaging Spectrometer \citep[LRIS;][]{oke95} and DEep Imaging Multi-Object Spectrograph \citep[DEIMOS;][]{fab03} on the Keck 10m telescopes. 

The early LRIS campaign targeted galaxies in the vicinity of two of the constituent clusters, Cl1604+4304 and Cl1604+4321 \citep[see][for further details]{oke95,gal04}, down to $i' < 23$. Data were taken using the 400 line mm$^{-1}$ grating blazed at 8500\AA~, with a central wavelength of 7000\AA~. This setup provides spectral coverage from 5500\AA~to 9500\AA, with a resolution of $\sim 7.8$\AA. The LRIS data were processed using standard IRAF tasks and scripts

The bulk of the redshifts used in this study come from observation of 18 slitmasks with DEIMOS taken between 2003 May and 2010 June. The first 12 slitmasks targeted mainly galaxies with $20.5 \le i' \le 24$, with higher priority given to red galaxies based on their $r'-i'$ and $i'-z'$ colors. The remaining six slitmasks were designed to obtain a magnitude-limited sample to a depth of $F814W \sim 23.5$ across a $16.\arcmin7 \times 5\arcmin$ subsection of the field running roughly north to south covering clusters Cl1604+4304 and Cl1604+4321 (hereafter clusters B and D, adopting the naming convention of \citet{gal08}).

All DEIMOS slitmasks were observed with the 1200 l mm$^{-1}$ grating with a FWHM resolution of $\sim 1.7$\AA~ (68 km s$^{-1}$) and a typical wavelength coverage of 6385\AA~ to 9015\AA. The exposure frames for each DEIMOS slitmask were combined using a modified version of the DEEP2 \textit{spec2d} package \citep{dav03}. This package combines the individual exposures for each mask and performs wavelength calibration, cosmic ray removal and sky subtraction on a slit-by-slit basis, generating a processed two-dimensional spectrum for each slit. The \textit{spec2d} pipeline also generates a processed one-dimensional spectrum for each slit.

In this work, we use objects with high quality redshifts ($Q \ge 3$), measured from at least one secure feature and one marginally detected feature. We adopted the redshift range of the Cl1604 supercluster to be $ 0.84 < z < 0.96 $. Our spectroscopic catalog contains 2445 objects, of which 531 are supercluster members with $Q \ge 3$. 

\section{ANALYSIS}

\subsection{Photometry}

We first created a unified astrometric reference frame using {\it SWarp}, spanning the spatial extent of the ground-based and IRAC images. All of the ground-based images were then resampled onto this common astrometric system with a new pixel scale of 0.\arcsec174 pixel$^{-1}$, the smallest pixel scale of all the images. Regions with no data were filled with zeroes, and a flag map for each image was created to indicate regions of no data or bad pixels. We then coadded the resampled versions of the WFCAM $J$ and $K$, LFC $r'$, $i'$, and $z'$, and COSMIC $R$ and $i'$ pointings to produce an ultra-deep optical-NIR detection image. Photometry was performed by running SExtractor in dual-image mode, using the ultra-deep image as the detection map, from which objects were detected and photometric apertures were determined. We then applied the same photometric apertures to each single-band resampled image. Using the deep stacked images as the detection map has three merits. First, the combined images provide 
significantly improved detection ability for faint objects compared to each single band image. Second, this procedure ensures that photometry in each band is performed using apertures of the same size. Finally, the catalog produced in each band has, by construction, the same sources, so they are trivial to combine. 

The LFC $r'$, $i'$, $z'$ and COSMIC $R$ and $i'$ data were photometrically calibrated using SDSS. We refer the reader to \citet{gal08} for further details of the calibration process. For regions covered by overlapping pointings in the same filter, we use the measurement with lowest photometric error for our combined catalog.
The WFCAM $J$ and $K$ data were calibrated using the same procedure but with photometry defined by the Two Micron All Sky Survey \citep[2MASS;][]{skr06} and all magnitudes were converted to the \textit{AB} system using $J_{AB} = J_{2MASS} + 0.89384$, $K_{AB} = K_{2MASS} + 1.84024$. The final catalog was corrected for Galactic reddening using the dust map from \citet{sch98}.

Because the point spread function (PSF) of IRAC is significantly different from the ground-based images, we made a separate deep IRAC detection image by stacking the resampled IRAC ch1 and ch2 images and coadding them.  The photometry in the two IRAC bands was performed in a similar manner as for the ground-based images. The IRAC deep stacked image was used for object detection. We used a fixed aperture of 1.\arcsec9 for measurement, which is recommended by the SWIRE survey, and applied aperture corrections to account for the finite aperture size \citep{sur05b}.
MOPEX outputs IRAC images with flux densities in MJy/sr, which we convert to fluxes in $\mu$Jy using
\begin{equation}
 1 \mbox{ MJy/sr } = 8.46152 \mu \mbox{Jy}
\end{equation}
and from $\mu$Jy to AB magnitude with
\begin{equation}
 M_{AB} = -2.5 \log(\mu \mbox{Jy}) + 23.9.
\end{equation}
Finally we produced our master catalog by cross-matching the ground-based and IRAC catalog using simple nearest-neighbor matching. The offset limit was set to 1.\arcsec5 between the two catalogs. 

\subsection{Galaxy Stellar Masses from SED Fitting}
\label{stellar_mass}

We derived galaxy stellar masses for the full Cl1604 spectroscopic sample using the \textit{Le Phare} code \citep{arn99,ilb06}, based on $\chi^2$ fitting analysis of the spectral energy distributions (SED). The SED templates were generated with the stellar population synthesis package developed by \citet{bc03}. We adopted an exponentially decreasing star formation history (SFH), SFH $\propto e^{-t/\tau}$, with nine different value of decaying time scale $\tau$, ranging from $\tau = 0.1$ Gyr to $\tau = 30 $ Gyr. 
Three different metallicities, $0.2Z_{\odot}$, $0.4Z_{\odot}$ and $Z_{\odot}$ were used. We used a \citet{cal00} extinction law  with $E(B-V)$ = 0, 0.1, 0.2, 0.3, 0.4 and 0.5. 

We included photometry from $r', i', z', J, K_s$, and \textit{Spitzer} 3.6$\mu$m and 4.5$\mu$m, giving 7 total bands for our SED fitting. We required an object to be detected in at least 4 bands for SED fitting. 
The \textit{HST} ACS has significantly better angular resolution than ground-based telescopes and \textit{Spitzer}, and the photometric apertures in $F606W$ and $F814W$ are much smaller than those in other bands. Therefore these two bands were not used for our SED fitting. A total of 525 objects were included in our analysis.

The SED fitting was performed in two steps. First, all available ground-based and \textit{Spitzer} channel 1 and 2 photometric measurements for each object were used for fitting. After the initial fit, we rejected a photometric data point if it contributed more than 25\% of the total $\chi^2$ for that particular fit and deviated more than 10$\sigma$ from the model, and repeated the fit a second time with the remaining bands. This method assures that our SED fits are not affected by catastrophic mis-measured photometry with incorrectly measured small errors.  

\begin{figure}
\begin{center}
\includegraphics[]{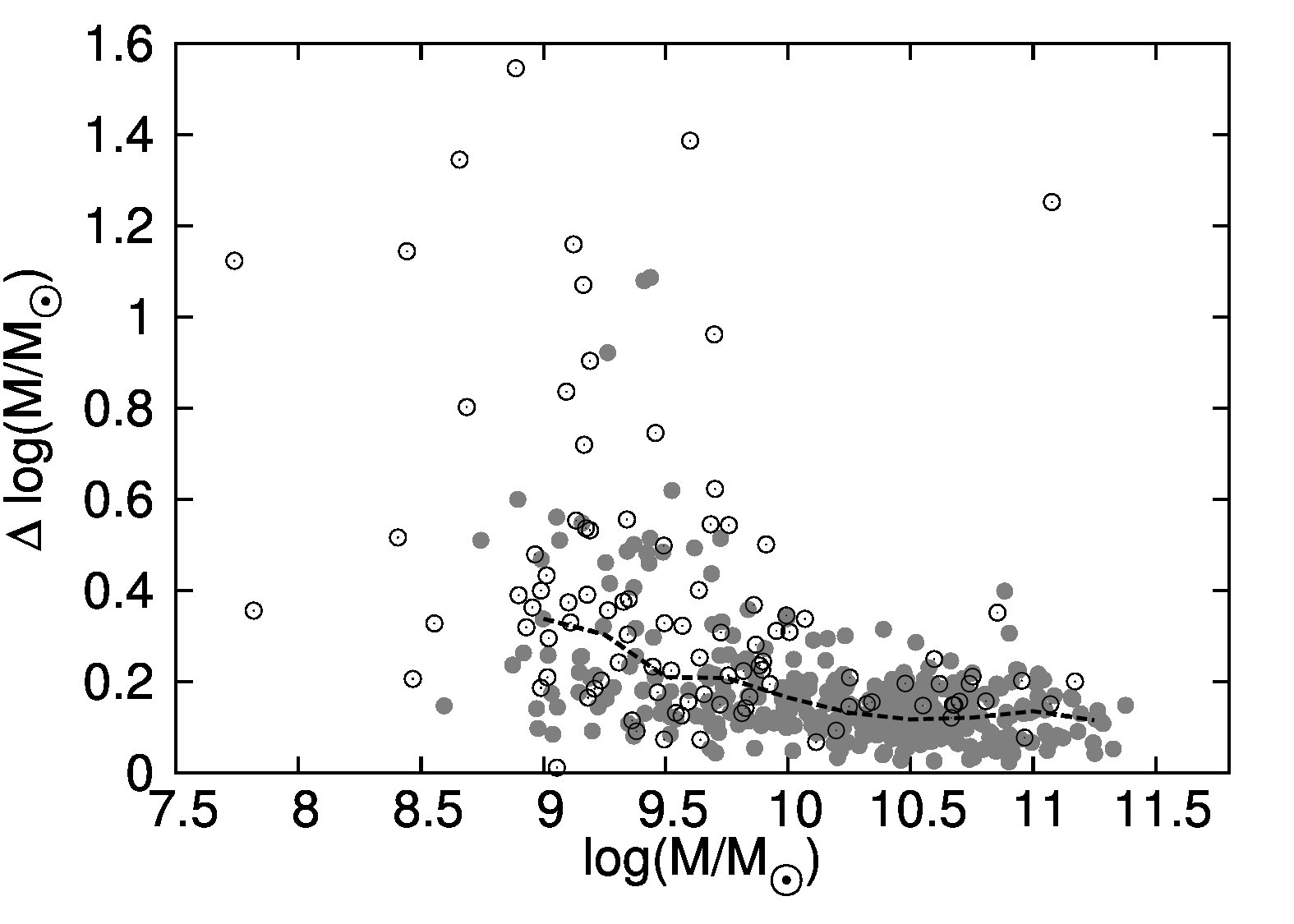}
\caption{Mass error as a function of galaxy stellar mass from SED fits. Galaxies detected in all 7 bands are shown as filled circles, while those detected in less than 7 bands but at least 4 bands are shown in open circles. The dashed line indicates the median mass uncertainty at a given stellar mass.} 
\label{fig:mdm}
\end{center}
\end{figure}

The uncertainty in the stellar mass of each galaxy is estimated by running 1000 Monte Carlo trials wherein the magnitude in each band is offset by a Gaussian random error, with the width of the Gaussian based on the measured photometric error of that object in that band. The standard deviation of stellar masses of 1000 trials is taken to be the stellar mass uncertainty. Figure~\ref{fig:mdm} plots the mass uncertainty versus stellar mass. Filled and open circles are objects detected in all 7 bands ($r', i', z', J, K_s, 3.6\mu m, 4.5\mu m$) and less than 7 bands, respectively. First, we find that, for a given mass, whether or not an object is detected in all 7 bands does not strongly affect the derived mass uncertainty. On the other hand, the mass uncertainty is mass dependent. For objects with stellar mass $< 10^{10} M_{\odot}$, the error increases significantly. 
The median and average $\Delta \log(M/M_{\odot})$ are 0.23 and 0.33, while for massive objects ($M > 10^{10} M_{\odot}$) they are 0.14 and 0.13, respectively. This is due to the less massive galaxies also being less luminous in the $K_s$, $3.6 \mu m$ and $4.5 \mu m$ bands and thus having correspondingly larger photometric errors.

Due to the typically poorer seeing in our imaging data compared to the spectroscopy, some of the spectroscopically identified galaxies are blended in ground-based images. Normally, this blending prevents us from obtaining correct photometry for these galaxies, and thus we cannot measure their stellar masses. Such objects were excluded from our analysis unless we could ensure that the blending did not significantly affect the photometry. First, if the object is separable in our \textit{HST} imaging, and our source of interest is brighter than the contaminating blended source by more than 2 magnitudes in both $F606W$ and $F814W$, we include it in our catalog. 
Second, if an object in the redshift range of interest blends with a background high redshift ($z>4$) spectroscopically identified object, we considered the effect of blending to be insignificant \citep[see][for an estimate of limiting magnitudes of high redshift galaxies in the Cl1604 field]{lem09}. After rejecting blended objects, we obtain stellar masses for 489 galaxies with $0.84 < z < 0.96$.

\subsection{K+A sample selection}
\label{sec:select}

Regardless of the physical trigger(s) that result(s) in a post-starburst galaxy, there are two basic components: (1) a significant star formation period in the recent past, followed by (2) no ongoing star formation. Practically, the [OII]3727 doublet emission line is often used as an indicator of ongoing star formation, and the H$\delta$ absorption feature as a proxy for the presence of A-type stars \citep[e.g.][]{pog99,dre99,oem09,ver10}. If one uses these spectroscopic diagnostics, a galaxy in its post-starburst phase should have no or little [OII] emission due to the lack of \ion{H}{2} regions that are forming O- and B-type stars, but have strong H$\delta$ absorption from the dominant A-type star population born in the recent past. 

While [OII] and H$\delta$ are usually used to select post-starburst samples, other diagnostics also have been adopted. For example, \citet{bla04} suggest using a combination of H$\delta$, H$\gamma$ and H$\beta$ equivalent widths (EW). A sample selected in this way can reduce contamination from dust obscured star-forming galaxies, in which the [OII] emission is highly attenuated. Instead of using Balmer line equivalent widths, \citet{qui04}, \citet{yan06} and \citet{yan09} decomposed galaxy spectra into young and old components, representing A- and K-type stars, respectively, and then used their ratio to indicate the relative stellar populations. 

We adopt the conventional criteria using limits on [OII] emission and H$\delta$ absorption EWs to be consistent with the majority of the literature in selection methodology. In the literature, various EW cuts have been adopted, depending on both data quality and on the desired contamination level. For example, \citet{dre99} and \citet{pog99} considered EW([OII]) $>-5$\AA~as ``absent``, \citet{ver10} adopted EW([OII]) $>-3$\AA~for the zCOSMOS survey, and \citet{got05} applied a more stringent cut of EW([OII]) $>-2.5$\AA~for SDSS. This inconsistency among studies leads to difficulty in comparing results. 
In this paper, we define two samples, both with EW(H$\delta$) $>$ 3\AA, but with different EW([OII]) limits, to examine the effects of different selection criteria. The O5 sample consists of galaxies satisfying the EW(H$\delta$) limit and with EW([OII]) $>-5$\AA. The more stringent O3 sample contains galaxies meeting the EW(H$\delta$) limit and with EW([OII]) $>-3$\AA. For discussion purposes, we further define an intermediate O35 sample as galaxies belong to the O5 sample but not the O3 sample, i.e., galaxies with EW(H$\delta$) $>$ 3\AA~and $-5$\AA~$<$ EW([OII]) $<-3$\AA. Errors in the EW measurement of our K+A samples are typically 1 to 3\AA. We will discuss this effect on sample selection in Appendix~\ref{sec:err}. Out of 489 galaxies with stellar masses, 48 galaxies belong to the O5 sample and 31 galaxies belong to the O3 sample. In this paper, if not further specified, the term ``K+A galaxy`` refers to the larger O5 sample.

We estimate the SFR corresponding to our adopted EW([OII]) limit using the conversion between [OII] luminosity, L([OII]), and SFR of \citet{kew04}. To convert EW([OII]) to L([OII]), we use the $i'$-band magnitude to estimate the continuum as it covers the wavelength of [OII] emission at $z \sim 0.9$. For a galaxy at $z=0.9$ (the median redshift of the Cl1604 supercluster), $i'=22.2$ (the median $i'$ magnitude of the O5 sample) and assuming $E(B-V) = 0.3$ \citep{lem12}, EW([OII]) = $-5$\AA~and $-3$\AA~correspond to SFR of $\sim 1.3$M$_{\odot}$ yr$^{-1}$ and $\sim 0.8$M$_{\odot}$ yr$^{-1}$, respectively.

\subsection{Cluster and Group Membership}
To examine the large scale environmental effects on K+A galaxies, we attempt to assign each spectroscopically confirmed member to one of the eight member clusters and groups in the Cl1604 supercluster, using both the galaxy's position and velocity offset relative to the central position and redshift of each cluster/group. We associated a galaxy with a specific cluster or group if (1) its velocity offset from the systemic velocity of the cluster or group is $\delta_v < \pm 3 \sigma_v$, where $\sigma_v$ is the velocity dispersion of the system and (2) its projected distance to the cluster or group center $r_{proj} \le 2 R_{vir}$, where $R_{vir}$ is the virial radius. 

We determined $R_{vir}$ of each cluster or group using the relation $R_{vir} = R_{200}/1.14$ \citep{biv06,pog09}, where $R_{200}$ is the radius at which the mean density of a cluster or group is equal to 200 time the critical density of the universe at the corresponding redshift. $R_{200}$ is calculated by
\begin{equation}
 R_{200} = \frac{ \sqrt{3} \sigma_v }{ 10 H(z)},
\end{equation}
where $H(z)$ is the Hubble parameter at redshift $z$. 

\begin{figure}
 \begin{center}
 \includegraphics[]{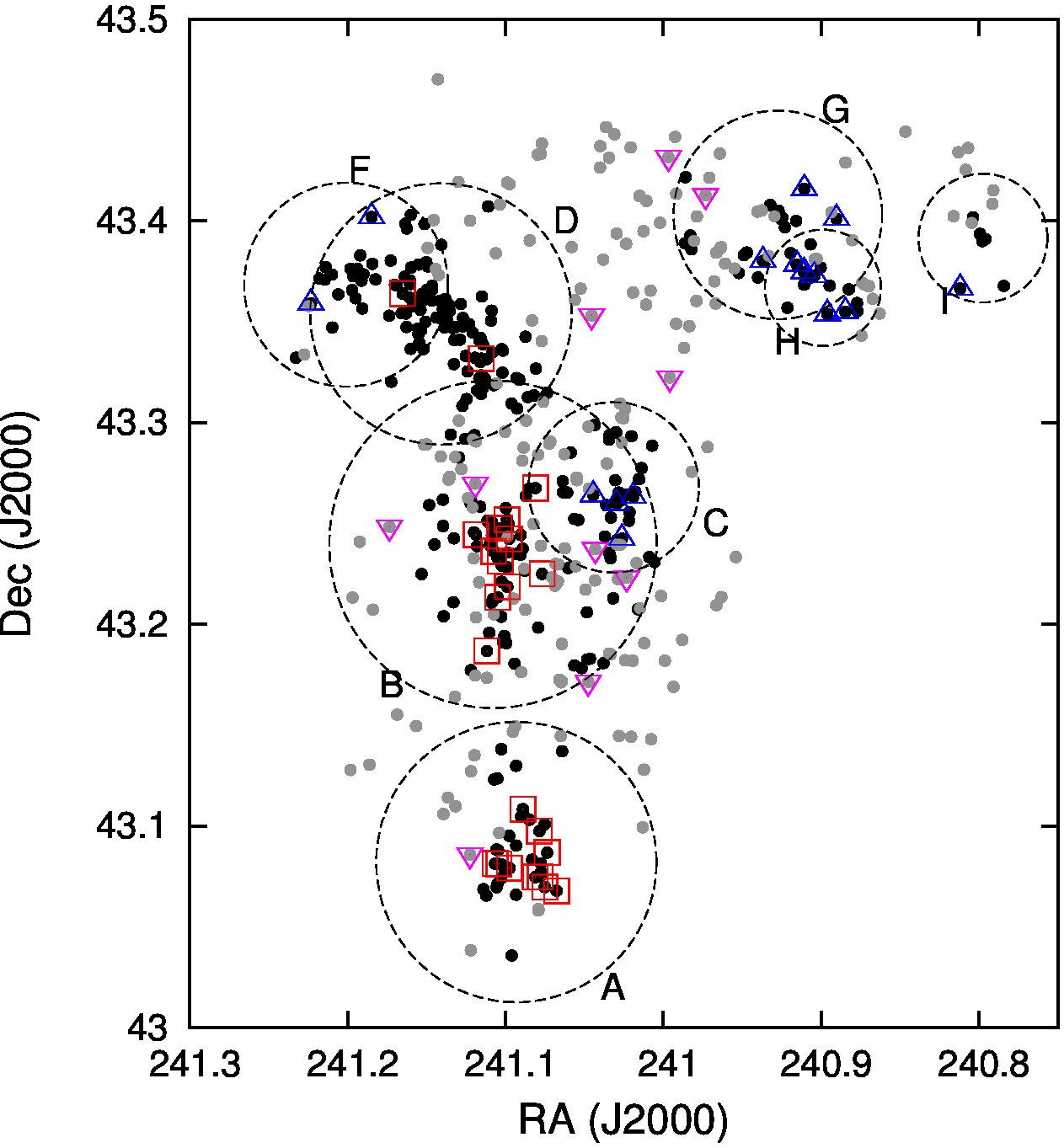}
\caption{Spatial distribution of K+A galaxies in the Cl1604 supercluster. Each dot represents a member of the Cl1604 supercluster. Black dots are cluster or group members, while gray dots are superfield members. squares (red), triangles (blue) and inverted triangles (magenta) indicate K+A galaxies in clusters, groups and the superfield, respectively. Each dashed circle is centered on a cluster or group, with a radius of $2 R_{vir}$ of the corresponding system. Most K+A galaxies in clusters (A, B and D) are located within $1 R_{vir}$ of their parent systems. As for group K+A galaxies, they are more evenly distributed. }
\label{fig:dist}
 \end{center}
\end{figure}

The spatial distribution of galaxies in the Cl1604 supercluster is shown in Figure~\ref{fig:dist}. Each black dot represents a galaxy associated with a specific group or cluster, while gray dots are superfield members. Dashed circles indicate a projected distance of 2 R$_{vir}$ from the center of each cluster/group. Squares, stars and crosses represent cluster, group and superfield K+A galaxies, respectively. Table~\ref{tab:sum} summarizes the properties of each cluster and group, including the RA, Dec, redshift, velocity dispersion, virial radius, total number of members, and number of members in the O5 and O3 K+A samples \citep{gal08,lem12,rum13}. In total, we have 489 galaxies with reliable stellar masses as well as [OII] and H$\delta$ EWs. We associated 297 galaxies in the field with a specific cluster or group in the Cl1604 complex. 
The remaining 192 galaxies fall within the redshift range of the supercluster but are found in filamentary structures near clusters or groups \citep{gal08}. We refer to these galaxies as the \textit{superfield} sample hereafter. 

\begin{table*}

\centering
\begin{threeparttable}
 \caption{Properties of the CL1604 Galaxy Clusters and Groups}
\begin{tabular}[t]{rcccccrrr}
\hline
\hline
Name & $\alpha_{J2000}$ & $\delta_{J2000}$ & $\bar{z}$ & $\sigma_v$ & $R_{vir}$ & $N$ & $N_{O5}$ & $N_{O3}$ \\
     &                  &                  &     &  (km s$^{-1}$) & $h_{70}^{-1}$Mpc & & &\\
\hline
Cluster A & 241.09311 & 43.0821 & 0.8984 & 722.4$\pm$134.5 & 0.98$\pm$0.18 & 36 & 10 & 8 \\
Cluster B & 241.10796 & 43.2397 & 0.8648 & 818.4$\pm$ 74.2 & 1.13$\pm$0.01 & 82 & 11 & 8 \\
Group C & 241.03142 & 43.2679 & 0.9344 & 453.5$\pm$ 39.6 & 0.60$\pm$0.05 & 24 & 4 & 3 \\
Cluster D & 241.14094 & 43.3539 & 0.9227 & 688.2$\pm$ 88.1 & 0.92$\pm$0.12 & 91 & 2 & 2 \\
Group F & 241.20104 & 43.3684 & 0.9331 & 541.9$\pm$110.0 & 0.72$\pm$0.15 & 25 & 2 & 1 \\
Group G & 240.92745 & 43.4030 & 0.9019 & 539.3$\pm$124.0 & 0.73$\pm$0.17 & 23 & 4 & 2\\
Group H & 240.89890 & 43.3669 & 0.8528 & 287.0$\pm$ 68.3 & 0.40$\pm$0.10 & 10 & 4 & 2 \\
Group I & 240.79746 & 43.3915 & 0.9024 & 333.0$\pm$129.4 & 0.45$\pm$0.13 & 6 & 1 & 0 \\
Superfield & & & & & & 192 & 10 & 5 \\
Total & & & & & & 489 & 48 & 31 \\
\hline
\end{tabular}
\label{tab:sum}
\end{threeparttable}
\end{table*}

\subsection{Spectral Measurements}

\subsubsection{Equivalent Widths and $D_n$(4000)}

Equivalent widths (EW) of the [OII] and H$\delta$ features of each galaxy were measured using the bandpass method following \citet{lem10}. The bandpass measurement were performed by defining two ``continuum`` bandpasses, slightly blueward and redward of the spectral feature, which are used to estimate the stellar continuum across the emission feature. An additional ``feature`` bandpass is defined to encompass the spectral line. Any pixel with large variance values were removed from the continuum bandpasses. We do not remove similar pixels in the feature bandpass. The EW is defined as 
\begin{equation}
 \mbox{EW(\AA)} = \sum\limits_{i=0}^n \frac{F_i - C_i}{C_i} \Delta\lambda_{r,i},
\end{equation}
where $F_i$ is he flux in the $i$th pixel in the feature bandpass, $C_i$ is the continuum flux in the $i$th pixel over the same bandpass, and $\Delta \lambda_{r,i}$ is the rest-frame pixel scale of the spectrum (in \AA~pixel$^{-1}$). Uncertainties in the EW were derived using a combination of Poisson errors on the spectral feature and the covariance matrix of the linear continuum fit and are given by \citep{boh83}:
\begin{equation}
\label{eq:err}
 \sigma_{EW}(\mbox{\AA}) = \sqrt{\left( \sum\limits_{i=0}^n \frac{\sigma_{F,i}\Delta\lambda_i}{C_i} \right)^2 + \left( \sigma_C \sum\limits_{i=0}^n \frac{F_i\Delta\lambda_i}{C_i^2} \right)^2  }.
\end{equation}

We adopt the bandpasses for [OII] and H$\delta$ in \citet{fis98}, with modification by eye for each galaxy spectrum to avoid poorly subtracted airglow lines and to avoid contaminating features near the spectral lines of interest. A handful of galaxies were observed twice by DEIMOS in the survey campaign, with EW([OII]) measurable for 6 such galaxies while EW(H$\delta$) can be measured in 3 galaxies. For each galaxy, we measured the EWs and the uncertainties of each spectral line for both observations. We compare the difference of EW of the same spectral line measured from the two observations, $\Delta$EW, to the EW uncertainty, $\sigma_{EW}$. From the 9 total measurements, we find that the median value of $\Delta$EW$/ \sigma_{EW}$ is $\sim 1$, suggesting that the EW uncertainty in Equation ~\ref{eq:err} is a fairly good description of the actual EW uncertainty. 

We also measure EW(H$\delta$) and the $D_n$(4000) spectral indices of the composite spectra of our K+A samples to examine their average properties (see Section~\ref{sec:comp_spec}). The $D_n$(4000) is measured using the ratio of the blue (3850--3950 \AA) and red (4000-4100) continua as defined by \citet{bal99}. Mean flux density values are calculated from the $\sigma$-clipped spectrum of each region, with the $D_n$(4000) index defined as $D_n(4000) = \langle F_{\lambda,r} \rangle / \langle F_{\lambda,b} \rangle$. Errors on the $D_n$(4000) index are calculated from the variance spectrum in each region, again using $\sigma$-clipping to avoid regions of poor night sky subtraction or regions that fell within in the 10 \AA~CCD chip gap.

\subsubsection{Composite spectra}
\label{sec:comp_spec}

We generated composite spectra of K+A samples to examine their average properties. We generated composite spectra for 24$\mu$m-detected and undetected K+A galaxies separately because they may have distinct properties, as suggested by the star formation rates implied by their IR luminosities.

We made the composite spectra following the method of \citet{lem09}. Each spectrum was first normalized by the galaxy's total spectral flux prior to stacking, then coadded using variance weighting \citep[see][]{lem09}. We measured EW(H$\delta$), $D_n$(4000) and EW([OII]) in each composite spectrum and these are provided in Table~\ref{tab:spec}.

Because $D_n$(4000) is not impervious to dust, the $D_n$(4000) of 24$\mu$m-detected galaxies must be corrected for dust attenuation. For example, the $D_n$(4000) of a galaxy with an 0.5~Gyr simple stellar population (SSP) and $E(B-V) = 0.5$ would be $\sim 0.1$ larger than a galaxy with the same stellar population but no dust \citep{mac05}. To account for this effect, we estimate the average $E(B-V)$ by comparing the SFR derived from [OII] and $L_{TIR}$ by:

\begin{equation}
 \langle E(B-V) \rangle =  \frac{2.5\log( \frac{SFR(\langle L_{TIR} \rangle) }{ SFR(\langle L_{[OII]} \rangle) } ) } { k'_{[OII]} }
\end{equation}
where $k'_{[OII]}$ is the \citet{cal00} reddening curve evaluated at the wavelength of [OII].

We list the estimated $\langle E(B-V) \rangle$ in Table~\ref{tab:spec}. Given the $E(B-V)$ we derived, the de-reddened $D_n$(4000) of 24$\mu$m-detected galaxies would be up to $\sim 0.15$ and $\sim 0.10$ smaller than the measured values for the O5 and O3 samples, respectively. For 24$\mu$m-undetected galaxies, the $\langle E(B-V) \rangle$ cannot be derived in this why. We calculate an upper limit, using $\langle L_{TIR} \rangle < 3 \times 10^{10} L_{\odot}$. On the other hand, if we take the median $E(B-V)$ from the result of SED fitting, $E(B-V) = 0.1$, the effect is $\sim 0.02$ \citep{mac05}.  

\begin{table}[t]
\scriptsize
\centering
\begin{threeparttable}
\caption{Spectral Indices of Composite Spectra of K+A Galaxies}
 \begin{tabular}[t]{lccccc}
\hline
\hline
  Sample    &  EW(H$\delta$) & $D_n$(4000)\tnote{a} &  EW([OII])     & $<L_{TIR}>$           & $<E(B-V)>$ \\
            &     (\AA)      &                 &   (\AA)        & ($10^{10} L_{\odot}$) &          \\
\hline
  O5 non-IR &  4.12$\pm$0.17 & 1.594$\pm$0.007 & -0.61$\pm$0.20 & $< 3.0$                   & $<0.62$      \\ 
  O5 IR     &  5.27$\pm$0.19 & 1.406$\pm$0.006 & -3.41$\pm$0.20 & 21.5                  & 0.79     \\ 
  O3 non-IR &  3.86$\pm$0.18 & 1.610$\pm$0.008 & -0.41$\pm$0.21 & $< 3.0$                   & $<0.50$      \\ 
  O3 IR     &  6.22$\pm$0.33 & 1.454$\pm$0.009 & -1.48$\pm$0.35 & 12.4                  & 0.57     \\ 
\hline
 \end{tabular}
\begin{tablenotes}
 \item[a] Not corrected for extinction. See Section~\ref{sec:comp_spec} for details.
\end{tablenotes}

\label{tab:spec}
\end{threeparttable}
\end{table}

\subsection{Completeness Correction}
\label{sec:cmplt}

\begin{table*}
\centering
\begin{threeparttable}[h]
 \caption{Prevalence of K+A Galaxies}
\begin{tabular}[t]{lrrrrrrrrrrrrr}
 \hline
 \hline
System & N & O5 & O3 &  O5-IR   & O3-IR  &  $f^{uncorr}_{O5}$ & $f^{uncorr}_{O3}$ & $f^{uncorr}_{O5-IR}$ & $f^{uncorr}_{O5-IR}$ & $f^{corr}_{O5}$ &  $f^{corr}_{O3}$ &  $f^{corr}_{O5-IR}$ &  $f^{corr}_{O5-IR}$ \\
\hline
Cluster A    &  34 & 10 & 8  & 8  & 6  & 0.29 & 0.24 & 0.24 & 0.18 & 0.19 & 0.13 & 0.11 & 0.09 \\
Cluster B    &  73 & 11 & 8  & 7  & 6  & 0.15 & 0.11 & 0.10 & 0.08 & 0.12 & 0.08 & 0.06 & 0.05 \\ 
Cluster D    &  80 & 2  & 2  & 1  & 1  & 0.03 & 0.03 & 0.01 & 0.01 & 0.05 & 0.03 & 0.02 & 0.02 \\ 
Groups       &  82 & 15 & 8  & 10 & 7  & 0.18 & 0.10 & 0.12 & 0.09 & 0.11 & 0.06 & 0.06 & 0.03 \\ 
Superfield   & 164 & 10 & 7  & 7  & 5  & 0.06 & 0.04 & 0.04 & 0.03 & 0.04 & 0.02 & 0.02 & 0.01 \\
All Galaxies & 437 & 48 & 33 & 33 & 25 & 0.11 & 0.08 & 0.08 & 0.06 & 0.07 & 0.04 & 0.04 & 0.02 \\
\hline
\end{tabular}
\begin{tablenotes}
 \item[NOTE] Only galaxies with $F814W < 24$, or $i'<24$ for those not covered by ACS, are included. The K+A fractions after completeness correction are derived as described in the text. Columns with "$-IR$" indicate K+A galaxies not detected at 24$\mu$m.

\end{tablenotes}
\label{tab:ks}
\end{threeparttable}
\end{table*}

\begin{figure*}
 \includegraphics[width=\textwidth]{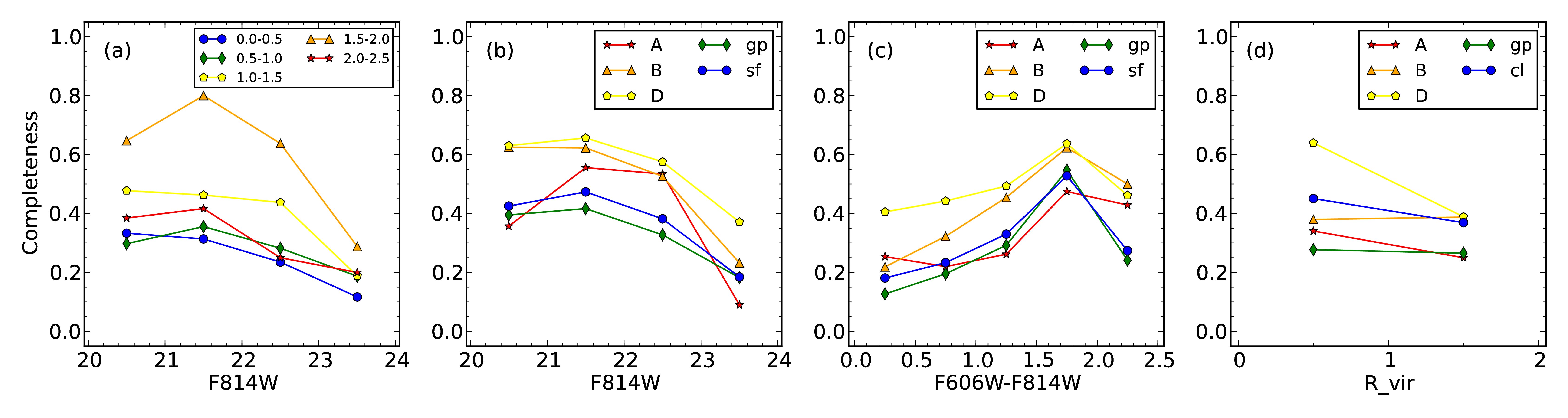} 
\caption{The spectroscopic completeness as a function of magnitude and color, environment, and distance to the cluster center. \textit{(a):} Completeness as a function of $F814W$ magnitude in each 0.5 mag $F606W-F814W$ color bin. \textit{(b):} Completeness as a function of $F814W$ magnitude, separated by environment. \textit{(c):} Completeness as a function of color, separated by environment. \textit{(d):} Completeness at different virial radii in clusters and groups. The three clusters are plotted separately, as well as a cluster composite. }
\label{fig:cmplt}
\end{figure*}

To determine the  true fraction of K+A galaxies, we estimated the spectroscopic incompleteness as a function of color and magnitude in each system in the Cl~1604, for galaxies with $F814W \le 24$, or $i' \le 24$ in regions not covered by ACS. For each system, we first plotted the CMDs of ACS photometric sources and all spectroscopic samples with ACS imaging in projection on the sky. The CMD was then separated into bins of 0.5 mag in color ($F606W - F814W$) and 1 mag in magnitude ($F814W$) over the color range $0 \le F606W - F814W \le 2.5$ and the magnitude range $20 \le F814W \le 24$. 
The completeness was calculated for each color-magnitude bin for a given system by dividing the number of objects with high quality redshifts by the number of photometric sources. Figure~\ref{fig:cmplt} shows the spectroscopic completeness dependence on magnitude, color, and host system. Also in Figure~\ref{fig:cmplt} we plot the radial dependence of the spectroscopic completeness in each cluster and the group composite. The completeness depends more strongly on a galaxy's location in  color-magnitude space than on its distance to the host system center. Therefore, we compute the color-magnitude dependent completeness in each system as whole, but do not further divide each system into smaller regions, resulting in more robust statistics for the dependence on  color and magnitude.

Assuming galaxies in the same color-magnitude bin have similar physical properties, that is, our spectroscopic samples are representative in each color-magnitude bin, the number of members in each color-magnitude bin was then corrected for incompleteness, and summing over all bins gave the estimate total number of members with ACS coverage in each system. However, the ACS imaging does not cover the full extent of the spectroscopically mapped areas of the supercluster. Therefore, to obtain a full estimate of the cluster membership, we must further correct for this incomplete ACS coverage. We apply a multiplicative factor to the ACS-based galaxy numbers, which is computed as the ratio of the total number of spectroscopic members to those with ACS coverage ($N_{spec}/N_{spec+ACS}$) within each structure. This correction factor is simply the inverse of the ACS sampling fraction in the region of each system. The K+A population was corrected in a similar manner. We first calculated the fraction of K+A galaxies 
in each color-magnitude bin. The number of K+A galaxies missed by our incomplete spectroscopy is then the number of missed system members multiplied by the K+A fraction in each bin. 

\begin{figure}
 \begin{center}
  \includegraphics[]{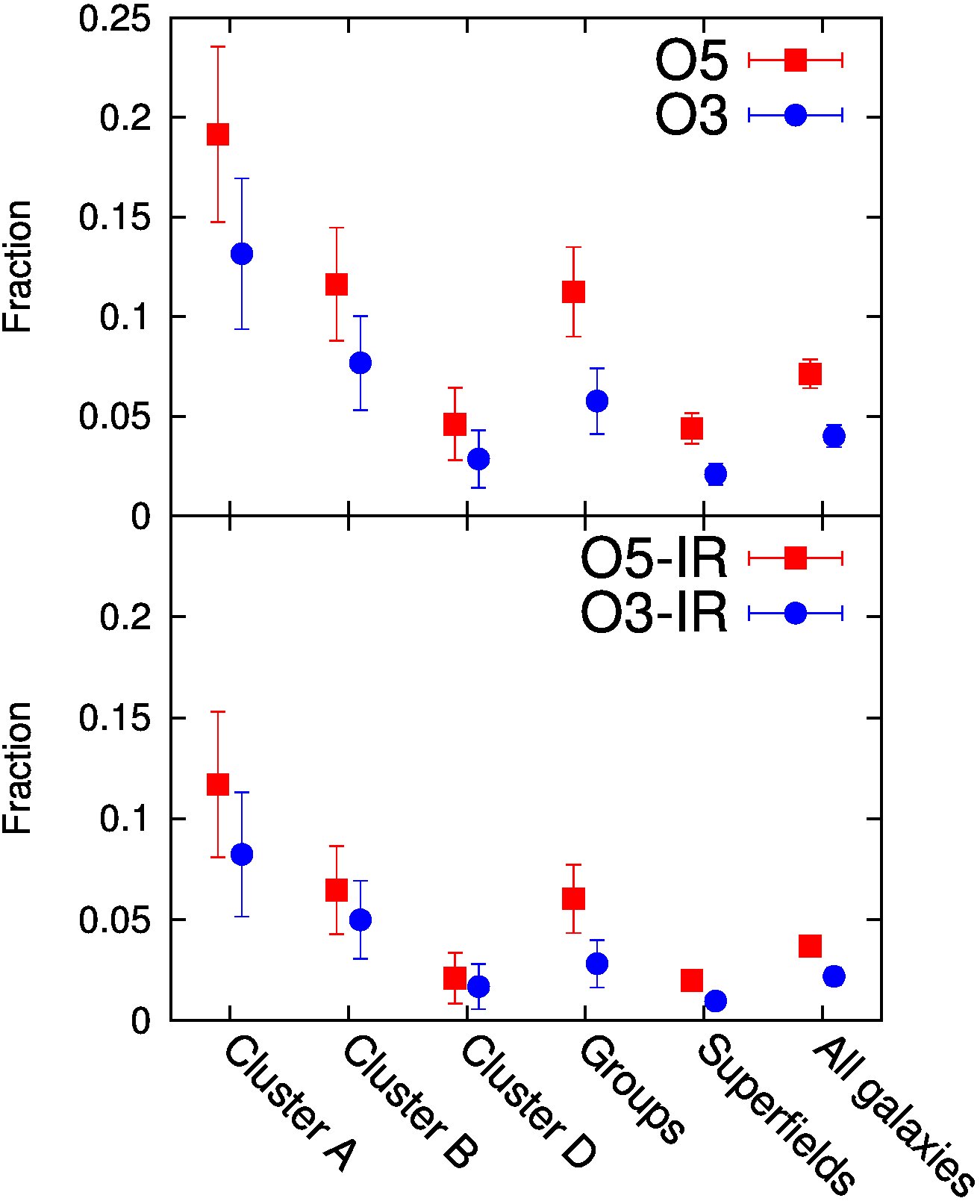}
\caption{Completeness-corrected K+A prevalence in the Cl1604 systems. The upper panel shows all K+A galaxies, and the lower panel shows only 24$\mu$m-undetected K+A galaxies. Squares (red) and circles (blue) are the O5 and O3 samples, respectively. The K+A prevalence correlates with dynamical stage of the cluster. The more evolved Cluster A has the highest K+A fraction, while dynamically young Cluster D has a low K+A fraction, comparable to that of the superfield.}
\label{fig:frac}
 \end{center}
\end{figure}

Table~\ref{tab:ks} lists the number of galaxies in each sample, as well as the completeness-corrected and uncorrected K+A fractions. The completeness-corrected K+A fraction in each system is plotted in Figure~\ref{fig:frac}. We also show the number and fraction of only 24$\mu$m-undetected K+A galaxies in Table~\ref{tab:ks} and Figure~\ref{fig:frac}. In this paper, all discussions of the K+A fractions will use the completeness-corrected values.

\subsection{Defining the Red Sequence}

We differentiated red and blue galaxies in the Cl1604 supercluster by using the \textit{HST} ACS color-magnitude diagrams (CMDs). The $F606W$ and $F814W$ passbands are effectively rest-frame U and B-bands at $z \sim 0.9$. The red sequence in each constituent system is determined by minimizing the $\chi^2$ of colors and magnitudes of member galaxies within a certain range to a linear model of the form
\begin{equation}
 F606W - F814W = y0 + m \times F814W.
\end{equation}
The detailed process is described in \citet{lem12}. Because of the small number of galaxies in each group, we combined all five groups into a composite. Table~\ref{tab:rs} summarizes the parameters of the red sequence fits for each system. For clusters, the red sequence width is defined by $\pm 3 \sigma$, while for the group composite and field samples, $\pm 2 \sigma$ is adopted because heterogeneous sub-samples are included across a variety of redshifts, resulting in a larger color dispersion. 

\begin{table}[b]
\centering
\begin{threeparttable}
 \caption{Red-Sequence Fitting Parameters}
\begin{tabular}[t]{lccr}
 \hline
 \hline
Name & Intercept & Slope & 1$\sigma$ Width \\
\hline
Cluster A & 2.20 & -0.020 & 0.046 \\
Cluster B & 3.24 & -0.065 & 0.048 \\
Cluster D & 3.21 & -0.062 & 0.045 \\
Groups & 2.95 & -0.051 & 0.076 \\
Superfields & 3.18 & -0.063 & 0.091 \\
\hline
\end{tabular}
\label{tab:rs}
\end{threeparttable}
\end{table}

\subsection{Morphology}
\label{sec:morp}
For galaxies covered by our ACS imaging, their morphologies were visually classified by one of the authors (L.M.L.). First, each galaxy is classified as either early-type (E or S0) or late-type (Sa through Sd and irregular). Second, galaxies showing tidal or disturbed features or ongoing mergers are further labeled as interacting. 
To estimate the reliability of our visual classification, a random subset of 150 supercluster members were presented to two of the authors (L.M.L. and R.R.G.) for classification. The result was then compared to the original classification. This process tests the consistency of visual classification from single observer and variation among different classifiers. We estimate roughly 5\%--10\% of our sample may be morphological misclassified \citep{lem12}.

\begin{figure*}
 \includegraphics[width=\textwidth,]{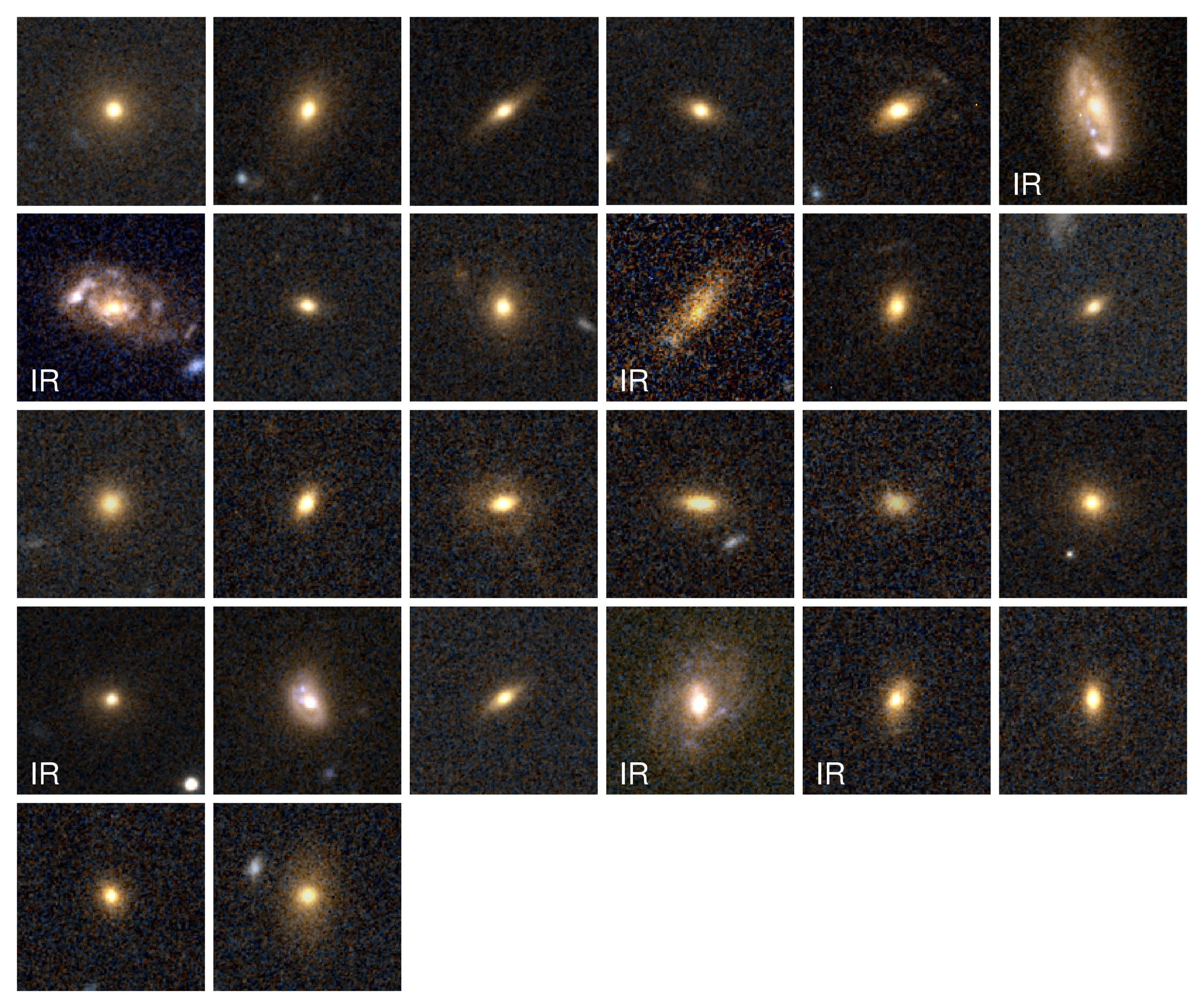}
\caption{Thumbnail images for the O3 sample. The color components are $F606W$ and $F814W$ for blue and red, respectively, while green is represented by a weighted average of the two. Each image is 3\arcsec~ across, equivalent to $\sim$23 kpc at the redshift of Cl~1604. Galaxies detected at 24$\mu$m are indicated in the images.}
\label{fig:o3}
\end{figure*}

\begin{figure*}
 \includegraphics[width=\textwidth,]{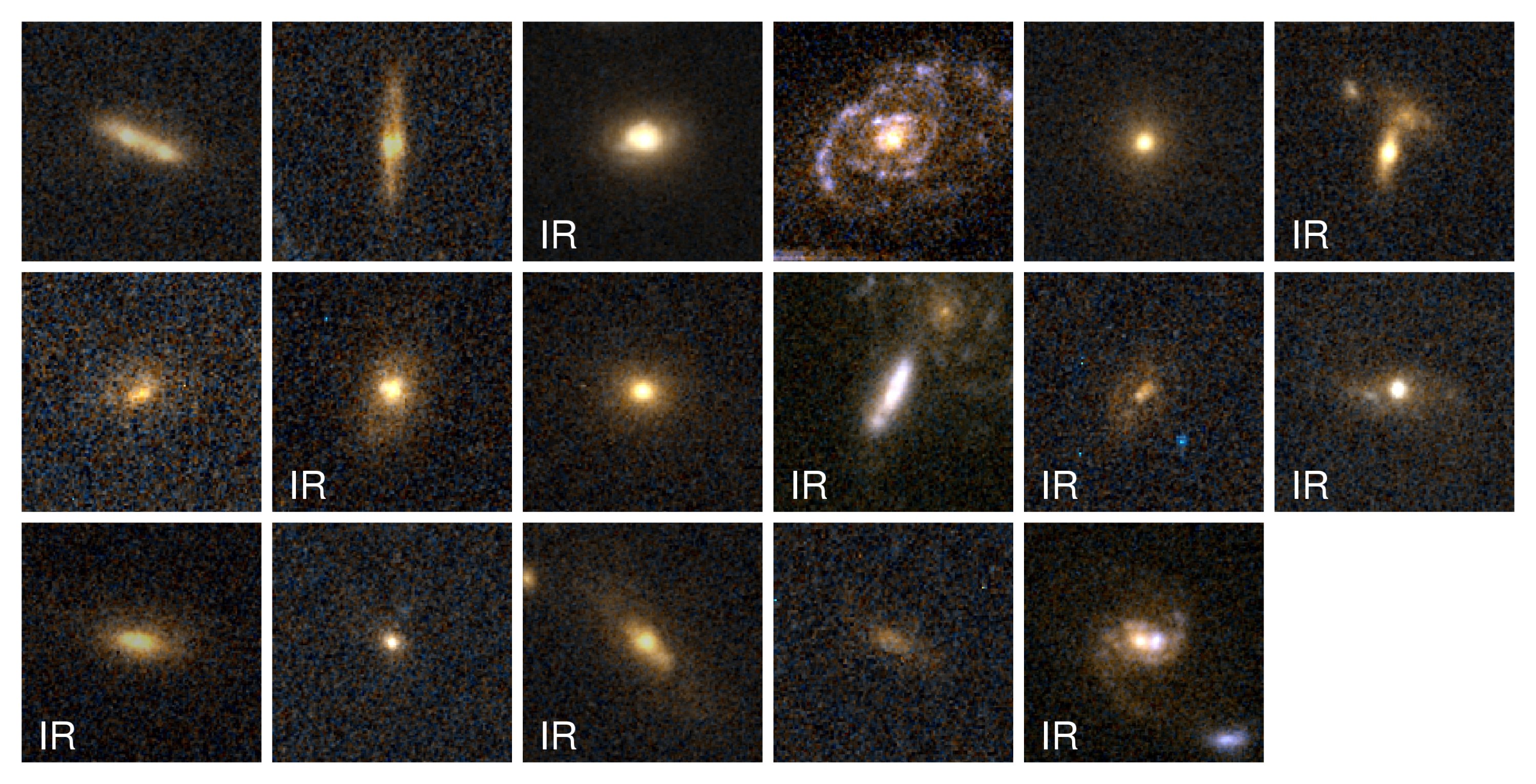}
\caption{Color thumbnail images for the O35 sample.}
\label{fig:o35}
\end{figure*}

We present thumbnail images of the O3 and O35 samples in Figures~\ref{fig:o3} and \ref{fig:o35}, respectively. Table~\ref{tab:mor} summarizes the statistics of the visual morphological classifications. Numbers in the parentheses indicate how many galaxies in each sample are also detected at 24$\mu$m. For both the more stringent O3 sample and the less restrictive O5 sample, K+A galaxies have heterogeneous morphologies. They are predominantly early-type galaxies and not interacting or merging with another galaxy, but late-type or interacting systems also constitute a non-negligible portion. We will discuss the morphological heterogeneity in Section ~\ref{sec:mech}.

\begin{table*}
\centering
\begin{threeparttable}
 \caption{Morphologies of K+A Galaxies}
\begin{tabular}{crrrrrrrrrrrrrrr}
 \hline
 \hline
Sample     & \multicolumn{3}{c}{O3} && \multicolumn{3}{c}{O35} && \multicolumn{3}{c}{O5} && \multicolumn{3}{c}{All Galaxies with HST} \\
\cline{2-4} \cline{6-8} \cline{10-12} \cline{14-16}
           &     Early & Late & All  &&   Early & Late & All &&   Early & Late & All  &&      Early & Late & All    \\
\cline{1-4} \cline{6-8} \cline{10-12} \cline{14-16}
Interaction&  0 (0) & 4 (3) & 4 (3)  &&  4 (4) & 5 (2) & 9 (6) &&  4 (4) & 9 (5) & 13 (9) &&   20 (9)  &  68 (32) & 88 (41)  \\
Non-inter. & 21 (2) & 1 (1) & 22 (3) &&  5 (2) & 3 (1) & 8 (3) && 26 (4) & 4 (2) & 30 (6) &&   171 (28) & 148 (53) & 319 (81)  \\
\cline{1-4} \cline{6-8} \cline{10-12} \cline{14-16}
All        &  21 (2) &  5 (4) & 26 (6) && 9 (6) &  8 (3) & 17 (9) && 30 (8) & 13 (7) & 43 (15) &&   191 (37) & 216 (85) & 407 (122)  \\
 \hline
\end{tabular}

\begin{tablenotes}
 \item \textbf{Note:} The numbers are different from those of Table~\ref{tab:sum} because the \textit{HST} ACS imaging does not cover the whole spectroscopic sample. Numbers of galaxies detected at 24$\mu$m are shown in parentheses.
\end{tablenotes}
\label{tab:mor}
\end{threeparttable}
\end{table*}

\section{PROPERTIES of K+A GALAXIES in the Cl~1604}

\subsection{Mass and Color}
\label{sec:prop}

\begin{figure*}
\includegraphics[width=\textwidth,]{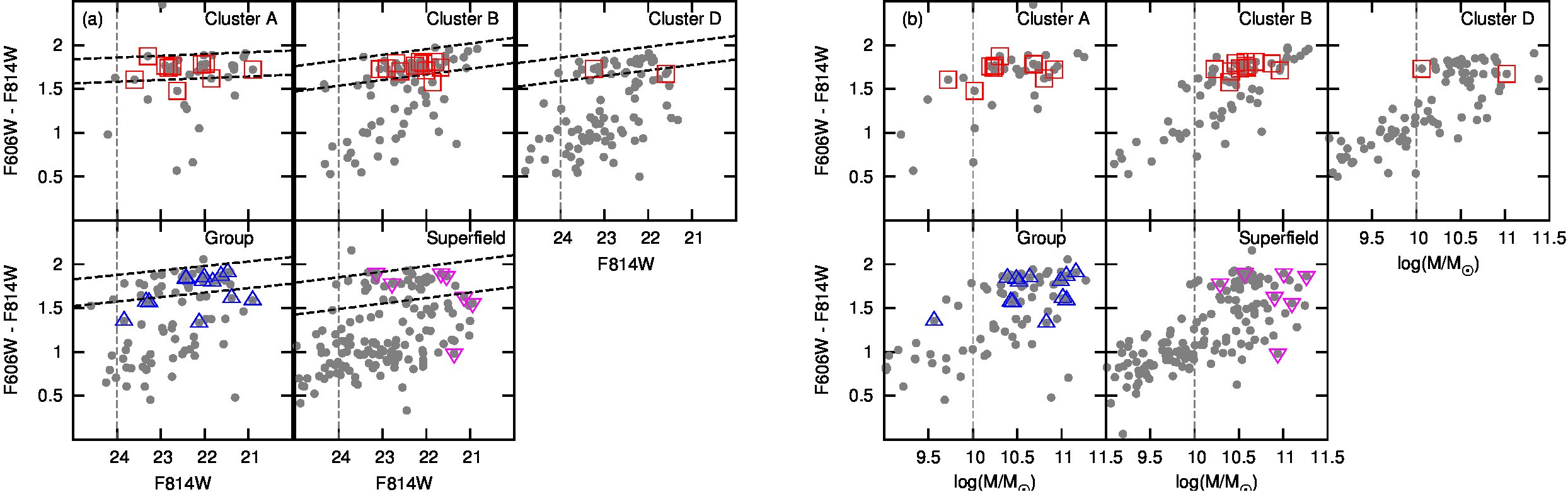}
\caption{(a) Color-Magnitude diagrams for galaxies in each cluster, the group composite and the superfield. The red sequence in each structure is indicated by the dashed lines. Each gray point represents a galaxy within the system. K+A galaxies are shown as squares (red), triangles (blue) and inverse triangles (magenta) for clusters, groups and superfield, respectively. The vertical dashed line at $F814W = 24$ represents the magnitude limit used in this paper. Only galaxies with $F814W < 24$ are included in discussion. (b) Color-stellar mass diagrams in each environment. Symbols are the same as in (a). The vertical dashed line at $\log(M/M_{\odot}) = 10$ labels our completeness limit on red-sequnce galaxies (see text).}
\label{fig:CMD}
\end{figure*}

\begin{figure*}
 \includegraphics[width=\textwidth,]{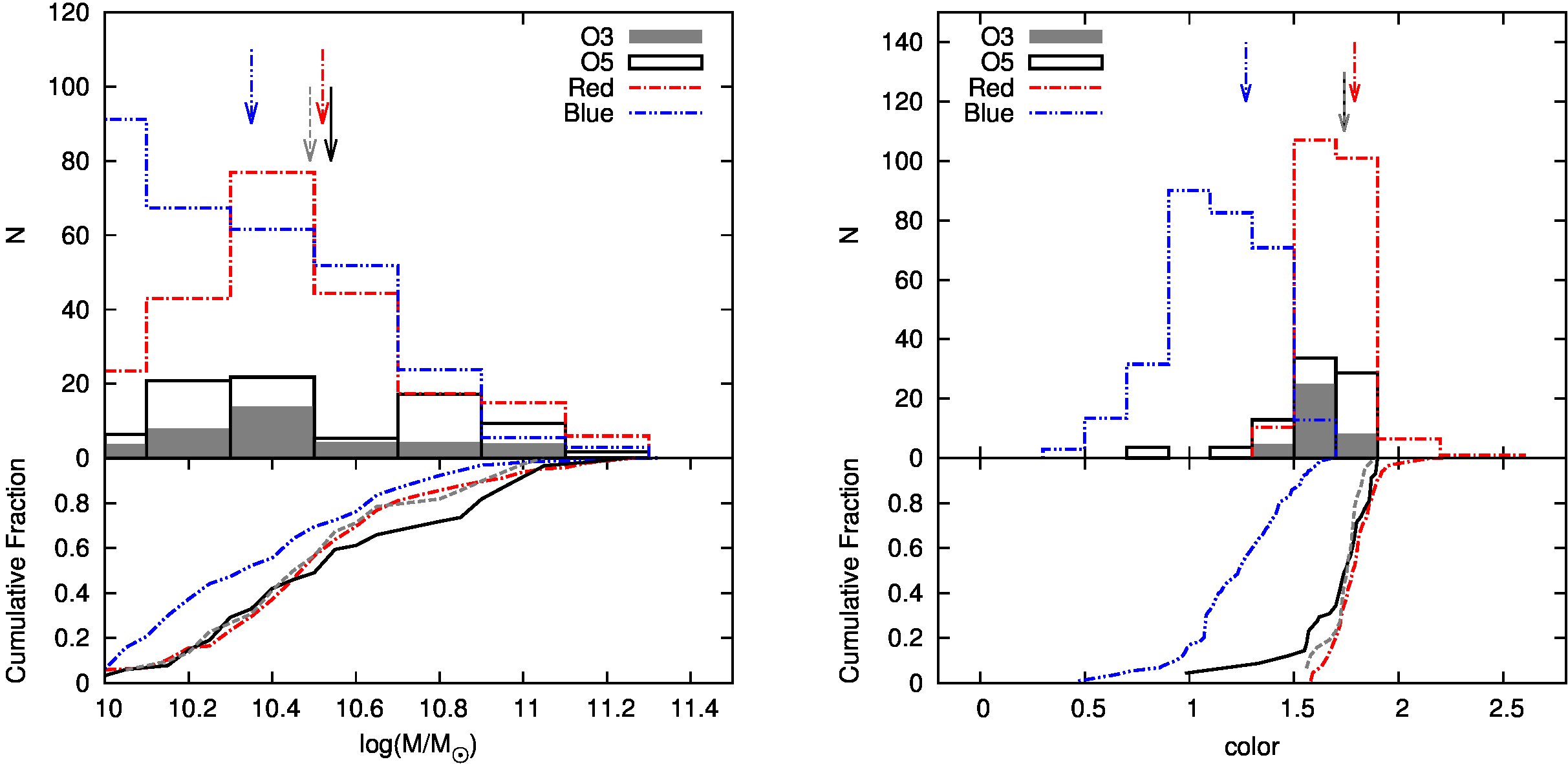}
\caption{Distribution of stellar masses and colors of red, blue, and K+A galaxies. The upper panels show the histograms. Red and blue galaxies are represented by dash-dotted (red) and dash-dot-dotted (blue) lines. The solid lines (black) indicate the O5 sample while the O3 sample is shown by the gray shaded regions. Arrows indicate the median stellar mass of each sample. The lower panels show the cumulative fraction of each sample, with gray dashed lines represent the O3 sample.}
\label{fig:hist}
\end{figure*}

Figure~\ref{fig:CMD} shows the color-magnitude diagram and color-stellar mass diagram of each cluster, the group composite and the superfield in the Cl~1604 supercluster. K+A galaxies are labeled with open symbols. In the color-magnitude diagrams, the red sequence is shown by the dashed lines. In Figure~\ref{fig:hist}, we plot the completeness-corrected color and stellar mass distributions of red, blue and K+A galaxies, respectively. Each galaxy is weighted by the inverse of the completeness calculated in Section~\ref{sec:cmplt}. For the stellar mass distribution, we only consider galaxies with $\log(M/M_{\odot}) \geq 10$ for two reasons. First, the uncertainty in stellar mass from SED fitting increases significantly below $\log(M/M_{\odot}) = 10$ (Figure~\ref{fig:mdm}). 
Second, for a passive galaxy with a near-instantaneous burst ($\tau = 0.1$ Gyr) star formation history, solar metallicity and formation redshift $z_f = 3$, $F814W = 24$ (the magnitude limit down to which we correct for the completeness) 
roughly corresponds to $\log(M/M_{\odot}) = 10$. This result means that red galaxies with $\log(M/M_{\odot}) < 10$ are excluded by the magnitude cut. Therefore, when examining stellar mass distributions, we only discuss galaxies more massive than $\log(M/M_{\odot}) = 10$.

We see that the majority of K+A galaxies lie in the red-sequence, with a minority in the blue cloud. Those K+A galaxies with bluest colors are in the O5 sample, possibly due to higher residual star formation rates. The median $F606W - F814W$ color of all K+A galaxies is 1.74, similar to the red sequence (1.79) and distinct from that of blue cloud galaxies (1.27) for galaxies with $\log(M/M_{\odot}) \geq 10$. K+A galaxies occupy the massive end of the mass spectrum, comparable to red-sequence galaxies. The massive nature of these K+A galaxies is consistent with those in the field at moderate to high redshifts \citep{ver10}. 
While \citet{ver10} argued that K+A galaxies with $0.48 < z < 1.2$ in the COSMOS field are on average more massive than red quiescent galaxies, we cannot distinguish the masses of K+A galaxies from red-sequence galaxies in Cl1604. The median stellar masses of red-sequence galaxies and K+A galaxies are $\log(M/M_{\odot}) = 10.54$ (10.49 for the O3 sample) and $\log(M/M_{\odot}) = 10.52$, respectively. Given the uncertainty of our stellar mass estimate, $\Delta \log(M/M_{\odot})$ = 0.14, the difference is insignificant. The similar colors and masses of K+A and red-sequence galaxies suggest that, at least in high-density regimes, K+A galaxies are likely the progenitors of some red-sequence galaxies, or vice-versa.

On the other hand, blue-cloud galaxies are less massive than K+A galaxies, with a median stellar mass of $\log(M/M_{\odot}) = 10.35$ and the number counts increasing to lower mass. The difference between the median stellar masses of K+A galaxies and blue-cloud galaxies is $\sim 10^{10} M_{\odot}$, about 50\%
of the stellar mass of an ``average`` blue-cloud galaxy. In our sample, a K+A galaxy has a SFR less than $1 M_{\odot} \mbox{ yr}^{-1}$. Assuming that the SFR of an ``average`` blue-cloud galaxy drops to that of a K+A galaxy immediately and retains SFR $\sim 1 M_{\odot} \mbox{yr}^{-1}$ for 1~Gyrs, the stellar mass only increases by $\sim 10^9 M_{\odot}$ and thus cannot reach the mass of an ``average`` K+A galaxy passively. This result suggests that, for an average blue-cloud galaxy turning into a K+A galaxy, there should be an epoch of rapid mass assembly before the SF shuts down. Further discussion will be presented in Section~\ref{sec:mech}

\subsection{The Environment}

The Cl~1604 supercluster contains galaxies in a variety of clusters and groups, with velocity dispersions ranging from $\sim 300 - 800$ km s$^{-1}$, as well as those not belonging to any bound structure, providing a variety of environments all at similar redshifts. The prevalence of K+A galaxies range from 4\% in the superfield to 19\% in Cluster A. Even the three clusters, which have similar velocity dispersions, and thus possibly similar mass, do not have similar K+A fractions. This variation among individual clusters has been shown in several previous studies \citep{dre99,tra03,pog09}. 
One source of this variation may be the cluster mass, as \citet{pog09} found from 20 intermediate redshift clusters in the ESO Distant Cluster Survey (EDisCS). They found that the K+A fraction correlates with the velocity dispersion of the host cluster, after binning their sample into three velocity dispersion bins. However, in each bin, the variation in the K+A fraction is still large. This variation is also seen in  Cl~1604, where the three clusters have quite similar velocity dispersions, but in Cluster A, K+A galaxies are almost 4 times more prevalent than in Cluster D, whose K+A fraction is comparable to the superfield. Although the dependence on cluster mass may be present, it cannot be the only factor. The dynamical state of each cluster may also be important. 

\begin{figure*}
 \begin{center}
  \includegraphics[width=\textwidth,]{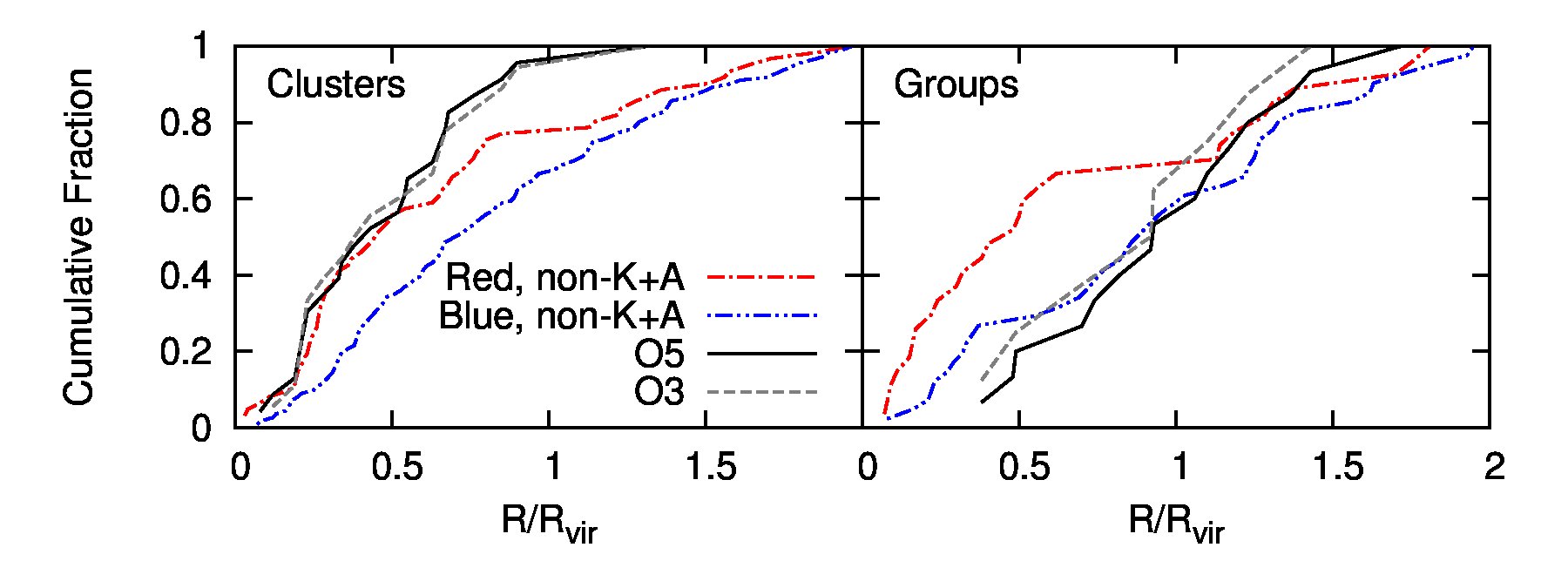}
\caption{Cumulative projected radial distributions of different types of galaxies as a function of virial radius. The projected distance is first normalized by the virial radius of each cluster or group, and then combined. In clusters, K+A galaxies are centrally concentrated, even more than red-sequence galaxies. Group K+A galaxies, on the other hand, are evenly distributed.}
\label{fig:cumu}
 \end{center}
\end{figure*}

In addition to the prevalence of K+A galaxies, we also examined the spatial distributions of K+A galaxies in the clusters and groups. Figure~\ref{fig:cumu} shows the cumulative projected radial distributions of K+A galaxies, along with those of red and blue galaxies, in clusters and groups as a function of virial radius. We combine clusters and groups by normalizing the projected distance of each galaxy by the virial radius of its parent system. In clusters, K+A galaxies are almost all located at $< 1 R_{vir}$ from the center but not in the infalling region of $1 < R_{vir} < 2$, while K+A galaxies in the lower-mass groups can be found at all radii. This distribution is unlikely an artifact of our spectroscopic sampling, as the average completenesses in clusters and groups does not strongly depend on the projected distance to system centers (Figure~\ref{fig:cmplt}). 
Thus the distribution of K+A galaxies may be a clue to the preferred locations where  quenching occurs in clusters and groups. Both the variation in K+A prevalence and radial distribution will be discussed in Section~\ref{sec:dis}

\section{DISCUSSION}
\label{sec:dis}

\subsection{K+A Galaxies and Host Cluster Dynamical State}
\label{sec:evol}

The three clusters in the Cl1604 supercluster are in different dynamical states, resulting in a range of physical properties \citep{gal08,koc09,lem12}. Among the three clusters, Cluster A is the most X-ray luminous and most massive, and shows clear velocity segregation between red and blue galaxies \citep{gal08,koc09}. It likely formed at an earlier epoch so that its primordial galaxy population had more time to virialize and establish a different dispersion than the infalling galaxy population.
Cluster A also falls on the X-ray--optical scaling relation, the $\sigma_v$--$T$ relation \citep{koc09,rum13}, This relation arises if the ICM shares the same dynamics as the cluster galaxies, so that the ICM temperature scales with the cluster velocity dispersion as $\sigma_v \propto T^{1/2}$, further supporting  Cluster A as a virialized structure.
Cluster B has the second-highest X-ray luminosity but no velocity segregation. It also has been shown that the velocity dispersion of its member galaxies is higher than inferred from the temperature of the ICM \citep{koc09,rum13}. This excess in velocity dispersion is often interpreted as an un-virialized system, with high velocity infalling galaxies inflating the velocity dispersion. \citet{gal08} also noted that the redshift distribution of Cluster B shows evidence of a substructure or a triaxial system. These dynamics of its member galaxies indicate that Cluster B is not yet fully virialized and may still be undergoing collapse. 
Cluster D has the most spectroscopically confirmed members in the Cl1604 supercluster but is not detected in X-rays, suggesting that it has not yet developed a dense ICM. In addition, its members have a filamentary distribution \citep{gal08}, also indicative of an un-virialized system.

The X-ray and optical properties of these three clusters suggests that they are at different dynamical states of evolution, with Cluster A the most evolved and Cluster D the most dynamically young. This difference among their dynamical states also leaves imprints on their galaxies' star-formation properties.
\citet{koc11a} found increased star formation activity in Cluster D compared to Clusters A and B and to the field, where the fraction of cluster members detected at 24$\mu$m is higher in Cluster D than the other systems. From composite spectra of the galaxies in each system,  Cluster D members on average have the highest EW([OII]) and lowest $D_n$(4000), suggesting a higher SFR and younger stellar age \citep{lem12}. The elevated rate of 24$\mu$m sources is interpreted as enhanced star formation in clusters associated with actively infalling galaxies, with the likely trigger being interactions between galaxies \citep{koc11a}. This scenario is also supported by the radial distribution of 24$\mu$m-detected galaxies in clusters, which is less centrally concentrated than that of 24$\mu$m-undetected galaxies \citep[Figure~8 of][]{koc11a}.

Contrary to 24$\mu$m sources, the K+A fraction varies among clusters but with an inverse trend, with the most evolved Cluster A having a high K+A fraction and the the dynamically youngest Cluster D having few K+A galaxies. Figure~\ref{fig:frac} plots the K+A fraction in each system for our various sample selection criteria. This variation is the same in all sample selections and could reflect a real difference among clusters. 

As argued in \citet{koc11a}, starbursts induced by interactions and mergers between infalling galaxies are occurring in Cluster D, which results in the elevated 24$\mu$m source fraction seen in Cluster D. Meanwhile, we do not see the aftermath of these starbursts, the post-starburst galaxies, in Cluster D. The K+A fraction is comparable to that of the superfield, showing that in this early stage of cluster formation, where a cluster is actively accreting galaxies, starburst activity is triggered, but star-formation quenching is not yet enhanced. The enhancement occurs at a later stage of cluster assembly, when the starbursts fade and move into their post-starburst phase. Cluster B is likely an example of such an environment, where the K+A fraction increase to 11\%, about twice that of Cluster D and the superfield.

The high K+A fraction in Cluster A is intriguing. In the Cl1604 groups, 11\% of members are K+A galaxies, comparable to the fraction in Cluster B but much lower than the 19\% seen in Cluster A. If mergers and interactions are the only routes producing K+A galaxies, the higher K+A fraction in Cluster A implies a higher frequency of merger-induced starbursts in Cluster A than in the Cl1604 groups. The enhanced merger rate in cluster cores had been proposed by \citet{str06} and observed by \citet{oem09} in a cluster at $z \sim 0.4$. For this enhancement to happen, galaxies fall into the cluster as bound pairs or small groups rather than individually. The gravitational perturbation from the cluster to the orbit of the bound groups may cause the orbit to shrink, leading to a higher merging probability. 
If a substantial fraction of member galaxies of Cluster A were recently accreted via this route, it would explain the high K+A fraction. However, the relaxed dynamical state of Cluster A suggests against this scenario. Cluster A does not exhibit sub-structure that indicating a recent infall of a group of galaxies \citep{gal08}.

Alternatively, \citet{dre13} suggested that not every merger turns into a K+A galaxy, but passive galaxies merging specifically with a smaller star-forming galaxy can create K+A phases. The excess of K+A galaxies in Cluster A may arise from a distinct underlying galaxy population in Cluster A compared to the other systems. In Cluster A, red sequence galaxies comprise $\sim$60\% of members (after correcting for incompleteness), much higher than the $\sim$30\% in Cluster B, Cluster D and the groups, and $\sim$20\% in the field. The high fraction of red sequence galaxies appears to be in line with the scenario of \citet{dre13}, so that mergers in Cluster A are more likely to involve a passive galaxy as a major component, thus producing more K+A galaxies. For this assertion to hold, there should be more early-type K+A galaxies in Cluster A than in other systems in order to account for the excess of K+A galaxies. 
After examining morphologies of K+A galaxies in Cluster A, we found 5 early-types out of 9 K+A galaxies with ACS imaging. The early-type fraction of K+A galaxies in Cluster A is not higher than that in the whole Cl1604 supercluster (see Table~\ref{tab:mor}). Thus, the  presence of more passive galaxies in Cluster A does not lead to a higher K+A fraction, but the numbers are too small to draw a definitive conclusion. 

These results lead us to consider another possibility: interaction between galaxies and the ICM. Violent processes such as ram-pressure stripping could remove cold gas from a cluster galaxy in $\sim 10^7$ yr \citep{aba99}, producing a K+A spectrum. To remove all the gas of a Milky-Way-like galaxy, the ram pressure should meet the requirement \citep{gun72,tre03}
\begin{equation}
\begin{array}{lr}
\rho_{gas} v_i^2  >  2.1 \times 10^{-12} \mbox{ N m$^{-2}$} \\
                     \times \left(\frac{v_{rot}}{220 \mbox{ km s$^{-1}$}} \right)^2 \left( \frac{r_h}{10 \mbox{ kpc}}\right)^{-1} \left( \frac{\Sigma_{\mbox{\ion{H}{1}} } }{8 \times 10^{20} m_H \mbox{ cm}^{-2}} \right)
\end{array}
\end{equation}
where $\rho_{gas}$ is the gas density, $v_i$ is the velocity of the galaxy and the rotational velocity $v_{rot}$, scale length $r_h$ and H$_{I}$ surface density of the galaxy in question are expressed in units for a Milky-Way-like galaxy. Making use of $Chandra$ X-ray data, the gas density profile of Clusters A and B can be derived from the best-fit isothermal $\beta$-models from \citet{rum13}. For a Milky-Way-like galaxy with $v_i = 800$ km s$^{-1}$, the stripping radii for both Clusters A and B are $\sim 0.3 R_{vir}$. 
That means ram-pressure stripping is effective only for galaxies passing very close to the cores of Clusters A or B. For Cluster B, still undergoing collapse, some of the members would be just accreting onto the cluster and not yet have passed close to the cluster core. Although the ICM in the core of Cluster B is dense enough to strip a galaxy's gas, not enough galaxies have traveled deep enough into the cluster core and experienced this effect. On the contrary, galaxies in the dynamically evolved Cluster A have had a higher chance to interact with the dense ICM because they had been in the cluster for a longer period of time. In this case, besides galaxy mergers, an extra mechanism kicks in, boosting the K+A fraction in Cluster A. 

The three clusters in Cl~1604 may be demonstrating one of the evolutionary tracks for galaxies assembling into clusters. At the early stage, a starburst may be triggered by a galaxy merger while a galaxy is infalling into the cluster. On average, the cluster has an actively star forming population, and quenching is not yet enhanced. 
At a later stage, those galaxies have past their starburst phase and turn into post-starburst galaxies, while new galaxies are still infalling into the cluster, with  starbursts being triggered. As the cluster evolves, fewer galaxies assemble into the cluster and it becomes virialized, with a dense ICM in the center. Earlier starbursts fade out, and the ram pressure becomes strong in the center, producing more post-starburst galaxies. However, this broad picture is based on only 3 clusters in the Cl~1604 supercluster. Future analysis of the full ORELSE survey, which has  similar quality  DEIMOS spectral coverage in 20 large-scale structures at $z \sim 1$, will be able to provide a more complete picture on the co-evolution of galaxies and clusters.

\subsection{More on Merger-induced K+A Galaxies}
\label{sec:mech}

\begin{figure*}[t]
  \includegraphics[width=\textwidth,]{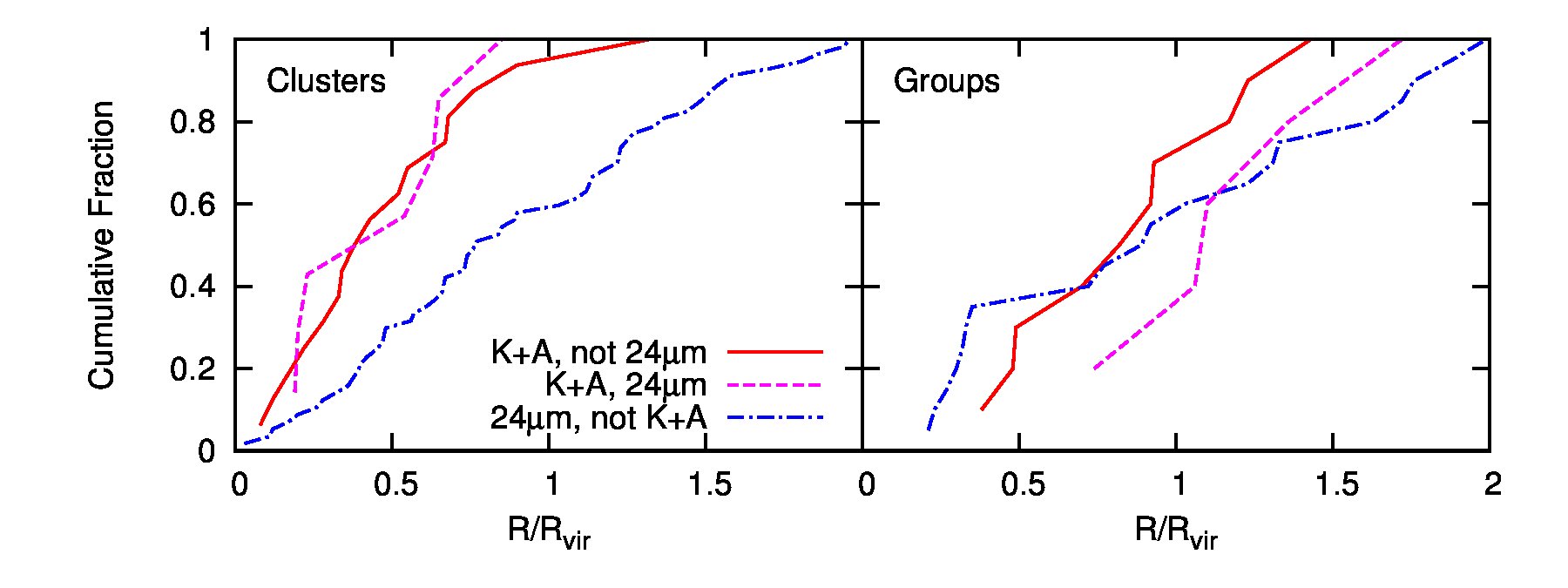}
\caption{Cumulative projected radial distributions of K+A and 24$\mu$m-detected galaxies in clusters and groups. 24$\mu$m-undetected K+A galaxies, 24$\mu$m-detected K+A galaxies, 24$\mu$m-detected non-K+A galaxies are represented by solid lines (red), dashed lines (magenta) and dash-dotted (blue) lines, respectively. In clusters, K+A galaxies share a similar projected radial distribution, regardless of whether they are detected at 24$\mu$m or not. On the other hand, the distribution of 24$\mu$m-detected K+A galaxies is distinct from other 24$\mu$m-detected galaxies. In groups, the three sub-samples show moderate differences.}
\label{fig:km_r_cum}
\end{figure*}

We have seen in Table~\ref{tab:mor} that about 30\% (13/43) of K+A galaxies exhibit interaction features. This simple morphological observation suggests that at least some K+A galaxies are related to galaxy mergers or interactions. Numerical simulations have demonstrated that galaxy mergers can result in star formation histories which produce K+A spectral features \citep{bek05,sny11}. A galaxy merger could trigger a short period of elevated star formation activity, followed by a rapid fall-off. A few hundred Myr after the peak of the starburst, the merger appears as a K+A galaxy, even before star formation activity fully stops. The whole K+A phase would last for another few hundred Myrs to as long as 1 Gyr, depending on the physical properties of the merging galaxies \citep{sny11}. 
On the other hand, disturbed morphological features fade away at a time scale comparable to that of the SFR decline \citep{lot08,lot10a,lot10b}. The merger-induced tidal arms can be prominent for $\lesssim 500$ Myr, but the low surface brightness tidal tails around the spheroid remnant are difficult to observe locally and nearly impossible at $z \gtrsim 0.5$ \citep{mih95,yan04}. It can be expected that at higher redshifts, $z \sim 0.9$, the observability of tidal arms decreases. Therefore, some merger-induced K+A galaxies may not be found to have interaction features, which means that the true population of merger-induced K+A galaxies could be much higher than the observed value of 30\% of the whole K+A population. Also with $HST$ images, \citet{ver10} reported a high incidence of asymmetry in K+A galaxies at $0.48 < z < 1.2$, supporting the hypothesis that K+A galaxies have experienced mergers or interactions in the recent past. 

The spatial distribution of K+As in clusters and groups in the Cl1604 supercluster is also consistent with a merger origin. In clusters, galaxy mergers are expected to happen more often at the outskirts than in the core, while in groups, the group centers are the preferred place for mergers. In the Cl1604 supercluster, the starburst activity distribution gives a consistent picture, where starbursts in clusters are found in infall regions and within $R_{vir}$ in groups \citep{koc11a}. So, for a starburst happening at the infall region of a cluster, the remnant likely ends up close to the cluster core after a few hundred Myr while infalling into the cluster. For a starburst galaxy near the core of a group, it can end up anywhere during the K+A phase, depending on whether it is moving towards or away from the core. 

Some of our K+A galaxies are bright at 24$\mu$m, implying that they still host substantial star formation. These galaxies do not meet the strict definition of post-starburst galaxies (no or little on-going star formation). Sometimes they are suggested to be dusty-starburst galaxies, and are therefore excluded from post-starburst galaxy samples \citep{oem09}. Figure~\ref{fig:km_r_cum} presents the cumulative projected radial distributions of 24$\mu$m-undetected K+A galaxies, 24$\mu$m-detected K+A galaxies and 24$\mu$m-detected non-K+A galaxies in clusters and groups. We found that the spatial distributions of K+A galaxies, whether they are detected at 24$\mu$m or not, are similar in clusters. 
We performed KS tests on the radial distributions of these different samples. In clusters, there is a strong suggestion (KS significance = 0.981) that the 24$\mu$m-detected K+A galaxies and their 24$\mu$m-undetected K+A counterparts are drawn from the same distribution. On the contrary, the 24$\mu$m-detected K+A galaxies and 24$\mu$m-detected non-K+A galaxies do not have similar radial distributions, with a KS likelihood of only 0.068 of being the same. The radial distributions in clusters suggest that 24$\mu$m-detected K+A galaxies are possibly associated with 24$\mu$m-undetected K+A galaxies but not with other 24$\mu$m-detected galaxies.  

Among all 24$\mu$m-detected K+A galaxies, the average IR luminosity for those with and without interaction features is $3.1 \times 10^{11} L_{\odot}$ and $1.3 \times 10^{11} L_{\odot}$, respectively. For 24$\mu$m-detected non-K+A galaxies, the average IR luminosity is $2.2 \times 10^{11} L_{\odot}$. The average IR luminosity of 24$\mu$m-detected, interacting K+A galaxies suggests elevated star formation activity in the interacting K+A galaxies, and the SFR drops as the interaction feature fades away. These 24$\mu$-detected K+A galaxies may be younger merger-induced K+A galaxies, whose SFR has not yet dropped to an undetectable level. This phase should last at most a few hundred Myrs. In Clusters A and B, it takes $\sim$ 1 Gyr for a galaxy with a velocity of 800 km s$^{-1}$ to traverse $\sim 1~R_{vir}$. Taking projection effects into account, the short-lived 24$\mu$m-detected K+A phase is hard to spatially distinguish from the parent K+A galaxy sample. The $D_n$(4000) indices and EW(H$\delta$) of composite 
24$\mu$m-detected and 24$\mu$m-undetected K+A galaxies (Table~\ref{tab:spec}) are also suggestive that both populations have truncated star-formation \citep{leb06,lem12}, where the 24$\mu$m-detected population is younger, indicated by its stronger EW(H$\delta$) and smaller $D_n$(4000).

Previous observations have also tested the idea that K+A galaxies are merger descendants. For example, \citet{wil09} compared K+A galaxies drawn from the VIMOS VLT DEEP Survey \citep[VVDS,][]{lef05} with numerical simulations and showed that K+A galaxies are consistent with being merger remnants that  underwent a strong starburst within the last few hundred Myr. Studies on the kinematics and spatial distributions of stellar populations in early-type K+A galaxies with integral field unit (IFU) spectroscopy also favors K+A galaxies as already coalesced gas-rich mergers \citep{pra09,pra12}. These studies  suggested that K+A galaxies are from gas-rich mergers, which induces likely strong starburst. 
However, \citet{dre13} claimed that passive galaxies are the principal progenitors of post-starburst galaxies, as they found the ratio between passive and post-starburst galaxies stays nearly constant across a wide range of environments from isolated galaxies to cluster cores at $z\sim0.5$, which indicates a tight correlation between these two populations. They suggested the dominant route of producing post-starburst galaxies is rejuvenated passive galaxies experiencing minor mergers or accretion. 

With our data, we are not able to directly measure the strength of the past starburst in the K+A galaxies, but their morphologies can give us a clue about the progenitor. In the Cl1604 supercluster, 13  of 43 K+A galaxies are late-type, about one-third of the population. Among the late-type K+As, 9  have interaction features. This simple observation shows that late-type galaxies constitute a non-negligible portion of the whole K+A population, but we can not rule out the route from rejuvenated passive galaxies for at least some of the population

The stellar mass of each population can give some more insight. In Section~\ref{sec:prop} we found that, on average, the stellar mass of a typical K+A galaxy is similar to that of a red-sequence galaxy, and  $\sim 50\% (\sim 10^{10} M_{\odot})$ more than the blue-cloud galaxy. This result means that, if an average blue-cloud galaxy turns into a K+A galaxy, its stellar mass should increase significantly in a short period of time. This increase can be achieved by a major merger but not by a minor merger. A major merger can produce a large amount of new stars rapidly, and the stellar mass of the merged galaxy is naturally nearly doubled by combining two progenitors with similar sizes. This scenario is only the case for blue-cloud K+A galaxies.  Because the stellar masses of red-sequence galaxies are similar to that of K+A galaxies, passive galaxies can experience a minor merger to create a K+A, and do not have to merge with another massive galaxy. 

\subsection{Post-starburst Galaxy Selection}

The [OII]$\lambda$3727,3729 line is a popular star formation indicator because it is accessible by optical spectroscopy from the local universe out to $z \sim 1.5$, allowing direct comparisons between samples over a wide range of cosmic time. However, [OII] emission has been shown to be a flawed star formation indicator. [OII] emission from LINER/Seyferts can be as strong as that in star-forming galaxies. Therefore, setting an upper limit on [OII] line strength tends to misidentify a LINER/Seyfert-harboring passive galaxy as a star-forming galaxy \citep{yan06,lem10,koc11b}. On the other hand, [OII] emission in a dusty star-forming galaxy would be highly attenuated, with the resulting EW([OII]) as low as that of a truly quiescent galaxy \citep{pog00}. These potential problems are often not addressed because there is a shortage of usable information to determine the sample incompleteness and contamination. For the Cl1604 supercluster, we are able to address these issues thanks to the extensive ancillary data. 

\subsubsection{LINER/Seyfert Activity}

Lacking LINER/Seyfert diagnostics for the whole spectroscopic sample, i.e., the ability to make a BPT or pseudo-BPT diagram \citep{bal81,sta06,jun11}, we estimated the number of LINER/Seyfert-harboring K+A galaxies based on a study of 17 [OII]-emitting absorption-line-dominated galaxies in Cl1604 presented in \citet{lem10}. Their absorption-line-dominated spectra indicate there is little ongoing star formation, so the [OII] emission may come from other ionizing sources. 
From near-IR spectroscopy, \citet{lem10} found about half of these [OII]-emitting absorption-line-dominated galaxies exhibit EW([OII])/EW(H$\alpha$) ratios higher than the typical observed value for star-forming galaxies. Furthermore, a majority of these galaxies have [NII]$\lambda$6584 to H$\alpha$ emission-line ratios consistent with a LINER/Seyfert origin, so the emission in many of these galaxies may be contributed by a LINER/Seyfert. Thus, these galaxies would have been categorized as K+A galaxies if not for the presence of a LINER/Seyfert. A strong color dependence is also observed: red [OII]-emitting, absorption-line-dominated galaxies more likely harbor LINER/Seyferts than blue ones.

We estimate the number of K+A galaxies that would be excluded due to [OII] emission from a LINER/Seyfert in the following manner.
From our spectroscopic sample, we first selected galaxies with (1) [OII] emission, (2) absorption features such as Ca H, Ca K and Balmer absorption lines, (3) no other emission features, such as Balmer lines, and (4) EW(H$\delta$) $\ge 3$\AA. These galaxies are candidate genuine post-starburst galaxies but were not classified as such because of [OII] emission produced by a LINER/Seyfert.  
From the 17 galaxies in \citet{lem10}, we calculated a color-dependent probability of an [OII]-emitting, absorption-line-dominated galaxy being LINER/Seyfert-harbored for each 0.5 mag color bin.
Applying this color-dependent probability, we derived the estimated number of LINER/Seyfert-harboring galaxies in our candidate sample. These galaxies with strong H$\delta$ absorption and LINER/Seyfert activity are possibly missing from our K+A sample because LINER/Seyferts contribute part of the [OII] emission. 

\begin{figure}
  \includegraphics[width=\columnwidth,]{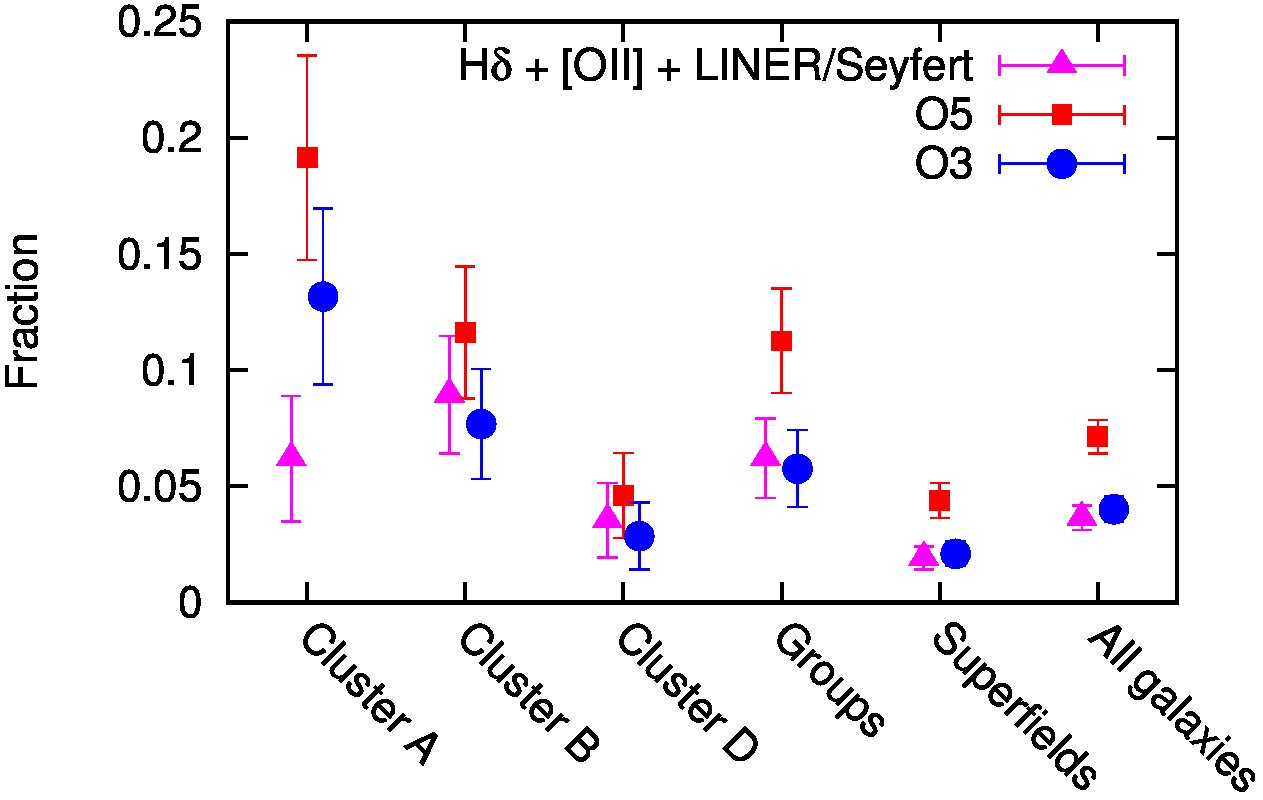}
\caption{Estimate of the fraction of K+A galaxies missed due to LINER/Seyfert activity. The triangles (magenta) represent fractions of LINER/Seyfert hosts with [OII] emission and H$\delta$ absorption in each system. The O5 and O3 samples are plotted in squares (red) and circles (blue) for comparison. Error bars are Poisson errors. Lacking additional diagnostics for LINER/Seyfert activity, the [OII]-selected K+A sample can suffer up to $\sim50$\% incompleteness in most of the systems.}

\label{fig:agn}
\end{figure}

Figure~\ref{fig:agn} plots the estimated numbers of these H$\delta$-strong, LINER/Seyfert-harboring galaxies along with the O5 and O3 samples in each system. Because we cannot quantify the amount of [OII] emission contributed from the LINER/Seyfert, this value should be taken as an upper limit on the number of missed K+A galaxies. In most of the systems, the number of LINER/Seyfert-harboring galaxies is comparable to that of the O3 sample and roughly half to two-thirds of the O5 sample. That is to say, K+A samples selected by means of EW([OII]) may suffer from incompleteness up to $\sim$50\%. Although it is derived from a limited sample size, this result at $z \sim 0.9$ is similar to what has been found for lower redshift SDSS galaxies \citep{yan06}. 

The exception is Cluster A, in which the number of LINER/Seyfert-harboring galaxies is only half of the O3 sample and one-third of the O5 sample. Because the estimate was made from a limited number of galaxies with LINER/Seyfert diagnostics, we cannot draw a firm conclusion whether or not this deviation is real. However, if Cluster A truly behaves differently from other systems, it suggests that the incompleteness of [OII]-selected K+A samples could depend on the properties of their parent systems. Thus, any comparison of the K+A prevalence among different samples should consider this possibility. For example, if we assume these LINER/Seyfert-harboring galaxies are all post-starburst galaxies, the total post-starburst fraction in Cluster A will still be higher than in Cluster B, but the difference becomes smaller. Future multi-object IR spectroscopy will be needed for post-starburst galaxy studies at high redshift in order to verify the effect of LINER/Seyferts and their possible environmental dependence. 

\subsubsection{Dust Obscuration}

If a galaxy has active on-going star formation and is dusty, its [OII] emission would be highly attenuated while still exhibiting strong H$\delta$ absorption, mimicking a post-starburst spectrum \citep{pog00}. As a result, post-starburst galaxy samples selected by means of EW([OII]) and EW(H$\delta$) are potentially contaminated by these dusty star-forming galaxies. With the \textit{HST} ACS and \textit{Spitzer} MIPS imaging, both morphology and mid-IR emission serve as extra star-formation indicators to test the effects of dust in Cl1604 members. 

With the assumption that early-type galaxies are passive and late-type galaxies are star-forming, we can check how many late-type galaxies were categorized as post-starburst galaxies, and thus estimate the contamination rate. In the O3 sample, 26 galaxies were imaged by ACS, with 21 early-types and 5 late-types. Nineteen percent of the O3 sample members are star-forming (late-type) galaxies. As for the O5 sample, 13 out of 43 members, or 30\%, are late-types. On the other hand, the infrared radiation provides another estimate of the star formation activity in these galaxies. The $L_{TIR}$ derived from the monochromatic 24$\mu$m flux generally matches well the values derived from multi-wavelength IR observations, with a scatter of $\sim$0.15 dex rms, but large deviations in individual sources can occur \citep{elb10}. 
Assuming the $L_{TIR}$ is properly recovered from the monochromatic 24$\mu$m flux, galaxies detected in our MIPS observation still have a SFR $> 5.2 M_{\odot}$ yr$^{-1}$. Our O3 and O5 samples have 6(19\%) and 15(31\%) galaxies detected at 24$\mu$m, respectively. Given that errors in both morphological classification and SFR derived from single band IR can exist, the morphology and mid-IR radiation give consistent estimates. Adopting the EW([OII]) $>$ -3\AA~limit, $\sim$ 20\% of K+A galaxies are still forming stars, and $\sim$ 30\% if the EW([OII]) $>$ -5\AA~limit is used. 

A closer look at the morphologies reveals that the majority of those star-forming (late-type or 24$\mu$m-bright) K+A galaxies have interaction features (see Table~\ref{tab:mor}). As discussed in Section~\ref{sec:mech}, these star-forming, interacting K+A galaxies likely have a decaying SFR and would turn into quiescent K+A galaxies in the near future. Considering their physical origin, including them in the sample is not necessarily contamination, but does increase the heterogeneity of the sample. 
Adopting different EW([OII]) limits can be thought of as accepting different levels of residual star-formation activity in a post-starburst galaxy. A stringent EW([OII]) limit selects mainly already coalesced mergers with little residual star formation, resulting in a more homogeneous sample, but at the cost of excluding objects from the same physical origin, i.e. pre-quenched merger-induced post-starburst galaxies, where the star formation is not completely halted. A loose EW([OII]) limit on the other hand give a more complete, but also more complex sample of merger-induced post-starburst galaxies with various morphologies and residual SFRs. 

Those galaxies having on-going star-formation features, K+A spectral type but no interaction features comprise roughly 10\% of the whole K+A sample. These galaxies are likely normal star-forming galaxies and are not related to post-starburst galaxies.

\section{SUMMARY}

The Cl1604 supercluster at $z\sim0.9$ consists of 8 clusters and groups, as well as galaxies in filamentary structures connecting the systems. This structure is the most extensively studied large-scale structure at such high redshift.
We selected our K+A galaxy sample using EW([OII]) and EW(H$\delta$). From all galaxies with EW(H$\delta$) $>$ 3\AA~, we selected two K+A galaxy samples with different EW([OII]) limits. The O3 and O5 samples consist of galaxies with EW([OII]) $>-3$\AA~and $-5$\AA, respectively. From 489 galaxies with measurements of both [OII] and H$\delta$ EWs and stellar masses, the O3 and O5 K+A samples contain 31 and 48 galaxies, respectively.

Combining these observations with \textit{HST} ACS and \textit{Spitzer} MIPS 24$\mu$m data, we find that:
\begin{enumerate}
 \item K+A galaxies in the Cl1604 supercluster mainly reside in the red sequence, with a minority located in the blue cloud. 
 \item K+A galaxies occupy the massive end of the mass distribution; they are as massive as other red-sequence galaxies. 
 \item We examined the projected clustocentric radii of K+A galaxies in clusters and groups. We find that K+A galaxies in clusters are centrally concentrated (even more than red sequence galaxies), whereas they are spread more widely in groups.
 \item K+A galaxies comprise $\sim 7\%$ of Cl1604 supercluster members, but the K+A fraction varies from 4\% to 19\% among the constituent clusters, groups, and the superfield. 
\end{enumerate}

From these results, combined with those from the literature, we conclude that:

\begin{enumerate}
 \item The K+A prevalence in clusters correlates with the dynamical state of the host cluster. At the early phase of cluster assembly, merger-induced starbursts occur more often in infalling galaxies. At this stage, we find more starburst galaxies but not K+A galaxies (Cluster D). When the merger-induced starburst fades away, the galaxy appears as a K+A galaxy, and the K+A fraction rises (Cluster B). As the cluster evolves, more galaxies have fallen deep into the cluster core, where star formation activity may be truncated due to interaction with the dense ICM, and the prevalence of K+A galaxies becomes even higher (Cluster A). 

 \item Interaction with a dense ICM can also produce K+A galaxies, but may be efficient only in dynamically evolved clusters.

 \item Major mergers of blue-cloud galaxies simultaneously explain the spatial distributions in clusters and groups, as well as the correlation between morphology and 24$\mu$m flux of K+A galaxies. The alternative scenario proposed by \citet{dre13}, that K+As arise from minor mergers involving a passive major component, is also possible, but cannot account for the whole sample.

 \item Because LINERs or Seyferts can also contribute [OII] emission, setting an upper limit on EW([OII]) to select post-starburst galaxies will miss some true post-starburst galaxies that harbor a LINER/Seyfert. The incompleteness can be as high as $\sim 50\%$ and may be environmentally-dependent, but we cannot draw a firm conclusion from our current data. 

 \item Roughly $30 \%$ of K+A galaxies selected by the EW([OII]) method are still forming stars. Adopting a more stringent EW([OII]) limit (EW([OII])$<$-3 \AA) yields a more homogeneous sample, $\sim 20\%$ of them have signs of star formation. The majority of these galaxies likely have a rapidly declining SFR and would become quiescent in the near future. About 10\% of the K+A sample are likely normal star-forming galaxies and are not related to post-starburst galaxies. 

\end{enumerate}

In this work, we studied the properties of post-starburst galaxies in the Cl1604 supercluster at $z \sim 0.9$, the most extensively investigated structure in the ORELSE survey. By studying galaxies at similar redshifts, we minimized the effect of cosmic evolution. However, study of a single structure may suffer from cosmic variance. The whole ORELSE survey covers a broader range of environment and dynamical states of clusters and groups, which will provide a more comprehensive picture of the cessation of star formation in high density regions at $z \sim 1$. 

\acknowledgments
We thank the anonymous referee for thorough comments and suggestions, which significantly improved this paper. Support for Program number HST-GO-11003 was provided by National Aeronautics and Space Administration (NASA) through a grant from the Space Telescope Science Institute, which is operated by the Association of Universities for Research in Astronomy, Incorporated, under NASA contract NAS 5-26555. We acknowledge support by the National Science Foundation under grant AST-0907858. This work is also based in part on observations made with the Spitzer Space Telescope, which is operated by the Jet Propulsion Laboratory, California Institute of Technology under a contract with NASA. Support for this work was provided by NASA through an award issued by JPL/Caltech. The spectrographic data presented herein were obtained at the W. M. Keck Observatory, which is operated as a scientific partnership among the California Institute of Technology, the University of California, and the National Aeronautics and 
Space Administration. 
The Observatory was made possible by the generous financial support of the W. M. Keck Foundation.
This publication makes use of data products from the Two Micron All Sky Survey, which is a joint project of the University of Massachusetts and the Infrared Processing and Analysis Center/California Institute of Technology, funded by the National Aeronautics and Space Administration and the National Science Foundation.
Near-IR data were taken with the United Kingdom Infrared Telescope, operated by the Joint Astronomy Centre on behalf of the Science and Technology Facilities Council of the U.K. As always, we thank the indigenous Hawaiian community for allowing us to be guests on their sacred mountain. We are most fortunate to be able to conduct observations from this site.

\clearpage

\appendix

\section{Effect of LRIS Spectroscopy on K+A Sample Selection}
\setcounter{figure}{0} \renewcommand{\thefigure}{A\arabic{figure}}
\setcounter{table}{0} \renewcommand{\thetable}{A\arabic{table}}
Because the spectral resolution of LRIS is relatively low (7.8 \AA), the K+A classification using LRIS spectra is more uncertain. In the Cl1604 field, some galaxies were observed by both LRIS and DEIMOS (Figure~\ref{fig:app}), and we use these galaxies to assess the effect of lower spectral resolution on identifying K+A galaxies. 

We examine issues of both contamination and completeness in the LRIS sample. First, we check how many LRIS-identified K+A galaxies are not classified as K+A from their DEIMOS spectra, that is, the cleanness of our LRIS K+A sample. Second, we examine how many DEIMOS K+A galaxies are not identified as such using the LRIS spectra, that is, the incompleteness. Figure~\ref{fig:app} compares EW measurements of [OII] and H$\delta$ from DEIMOS and LRIS. For both [OII] and H$\delta$, some emission lines measured to be moderately strong (EW $< -10 \AA$) from LRIS are seen as weak lines in DEIMOS. This result implies that some DEIMOS-identified K+As would be seen as emission line galaxies if they had only LRIS spectra and would thus be excluded from the sample. 

From 37 galaxies with both DEIMOS and LRIS spectra, we find 6 galaxies identified as K+As from their DEIMOS spectrum,  and none identified as K+As by the LRIS data. Thus, the cleanness of the LRIS K+A sample is not an issue - we do not see any galaxies identified as  K+A by LRIS but not by DEIMOS. In our full sample, the galaxies classified as K+A from LRIS constitute only a small fraction ($\sim10\%$) in our K+A samples: 6/48 in O5 and 3/31 in O3. Even if there is significant contamination by non-K+A galaxies in the LRIS-selected K+A sample, it will have little or no effect on our conclusions.

On the other hand, the incompleteness in the LRIS K+A sample could be significant. As noted above, 6 galaxies are identified as K+As by DEIMOS but none of them would be picked out by LRIS. Furthermore, if we assume that these duplicates are representative of LRIS galaxies in the Cl1604 spectroscopic sample, we would expect 13 K+As from the total of 82 LRIS galaxies if they had been observed by DEIMOS, whereas the LRIS spectra only pick out 6 and 3 in the O5 and O3 sample, respectively. We would like to point out that the duplicate lies mainly in the red sequence (Figure~\ref{fig:app}.(a)), where the K+A fraction is higher than average. Therefore, they may not be fully representative of all LRIS galaxies in Cl1604 and the expected total of 13  K+As is likely an overestimate. Nevertheless, this result suggests that we could be missing on the order of 5 K+As from our LRIS targets.

We perform Monte Carlo simulations to estimate the potential impact of missing K+As from the galaxies with LRIS spectra. In each of 1000 runs, we randomly select non-K+A LRIS galaxies and assign them as K+A galaxies, making the total number of LRIS K+A galaxies 13, then check their impact on the K+A fractions and spatial distributions.

Table~\ref{tab:lriska} lists the median, 16th, and 84th percentile of completeness-corrected K+A fractions from these 1000 realizations. Compared to Table~\ref{tab:ks}, the K+A fractions in groups are nearly unchanged because few group galaxies are LRIS galaxies. In clusters, the K+A fractions are mildly affected, but the relative K+A abundances in the three clusters do not change our interpretation of the results. Figure~\ref{fig:lriscum} shows the projected cumulative radial distributions of K+A galaxies in clusters and groups, similar to Figure~\ref{fig:cumu} but with the inclusion of the potentially missed K+As. The shaded region is the 16th and 84th percentile for the cumulative fraction at each radius. Again, the lower-resolution LRIS spectra have little to no effect on the results.

\begin{table}[h]
\centering
\begin{threeparttable}
 \caption{K+A prevalence after adding missing LRIS K+As}
\begin{tabular}[t]{lccccccccc}
 \hline
 \hline
Name &  \multicolumn{4}{c}{O5} & & \multicolumn{4}{c}{O3}  \\
\cline{2-5} \cline{7-10}
            & no LRIS & 16th & median & 84th & & no LRIS & 16th & median & 84th \\
\hline
Cluster A   & 0.19 & 0.20 & 0.23 & 0.26 & & 0.13 & 0.14 & 0.17 & 0.20 \\
Cluster B   & 0.12 & 0.14 & 0.15 & 0.17 & & 0.08 & 0.09 & 0.11 & 0.13 \\
Cluster D   & 0.05 & 0.07 & 0.09 & 0.11 & & 0.03 & 0.05 & 0.07 & 0.09 \\
Groups      & 0.11 & 0.11 & 0.11 & 0.12 & & 0.06 & 0.06 & 0.06 & 0.07 \\
Superfields & 0.04 & 0.04 & 0.05 & 0.06 & & 0.02 & 0.03 & 0.04 & 0.05 \\
\hline
\end{tabular}
\label{tab:lriska}
\end{threeparttable}
\end{table}

\normalsize

\begin{figure}
 \includegraphics[width=\columnwidth,]{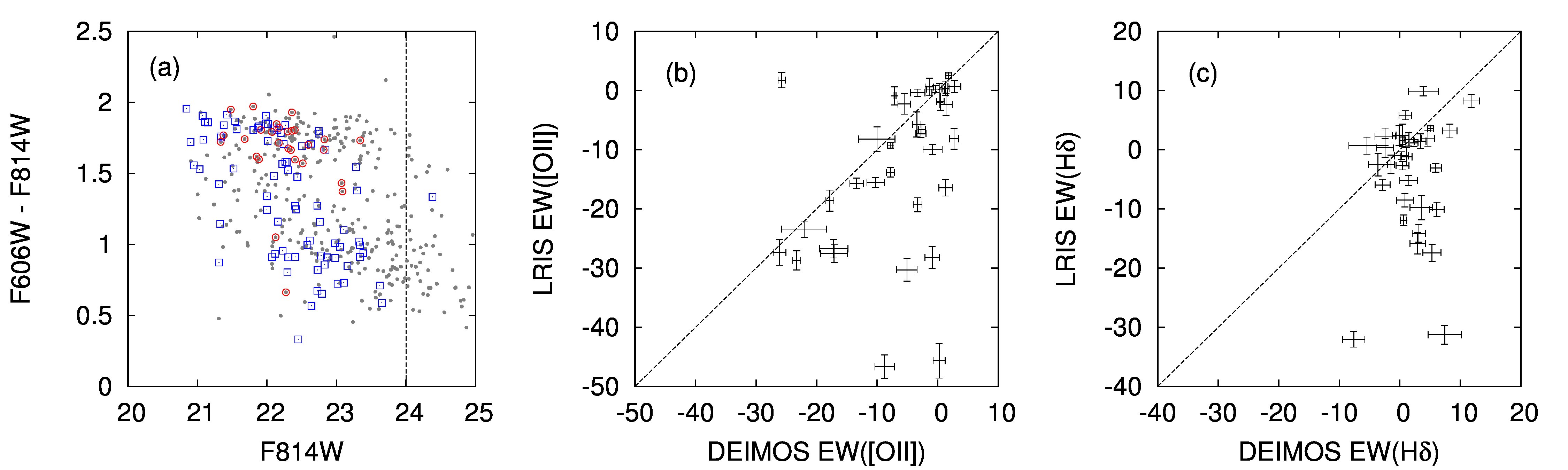}
\caption{(a) Color-magnitude diagram for the Cl1604 spectroscopic sample, separated by instrument. Galaxies with only DEIMOS spectra are plotted as filled circle (gray). Galaxies with only LRIS spectra are represented by open squares (blue). Galaxies observed by both DEIMOS and LRIS (duplicates) are shown as open circles (red). (b) Comparison between EW([OII]) of the same galaxy measured from DEIMOS and LRIS. (c) Comparison between EW(H$\delta$).}
\label{fig:app}
\end{figure}

\begin{figure}
 \includegraphics[width=\columnwidth,]{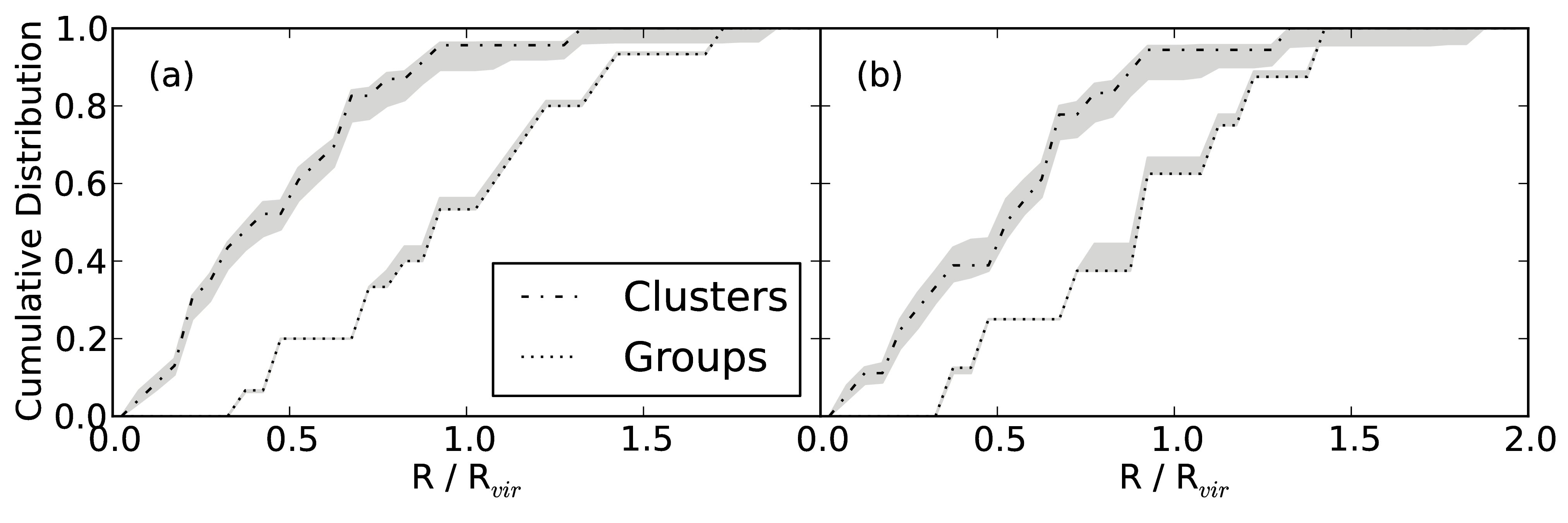}
\caption{Cumulative distributions of K+A galaxies in Cl1604 clusters and groups, for (a) the O5 sample and (b) the O3 sample, respectively. The dash-dotted and dotted lines show distributions without adding the potentially missing LRIS K+A galaxies, and shaded regions are the 16th and 84th percentiles of the distributions from 1000 Monte Carlo realizations that assign a certain number of LRIS non-K+A galaxies as K+A galaxies.}
\label{fig:lriscum}
\end{figure}

\section{Effect of EW Uncertainty on K+A Sample Selection}
\label{sec:err}

In this paper we discuss the K+A fractions in different environments, which have inherited uncertainties from the uncertainties of EW measurements. We test the robustness of the K+A counts using another Monte Carlo simulation. We run 1000 Monte Carlo trials wherein both the EW([OII]) and EW(H$\delta$) of each galaxy is offset by a Gaussian random error, with the width of the Gaussian based on the measured uncertainty of that spectral feature in that galaxy. 
We then apply the same K+A selection criteria to each simulated data set, obtaining the total number of K+A galaxies in each set. We then compute the standard deviation of numbers of K+A galaxies in each environment from the 1000 trials. 
The resulting uncertainty in the number of K+A galaxies is about 1.6 for Cluster A, Cluster B and the group composite, 0.8 for Cluster D, 2.1 for the superfield and 3.5 for the whole supercluster. The corresponding uncertainty in K+A fraction in each system is at level of $\sim 1--2\%$, except for the $\sim 4\%$ in Cluster A. Our conclusion is largely unaffected, except that the excess in Cluster would be less significant.

\section{The Effect of Sky Lines on Spectral Measurements}

Bright sky lines that are not properly subtracted can potentially affect spectral measurements. At the redshifts of the Cl1604 galaxies, the telluric A band at $7595-7670 \AA$ overlaps the continuum range used for measuring $D_n$(4000), and the bandpasses used for EW(H$\delta$) at 4102 \AA. We use the same ``bandpass'' method applied to the EW measurements to estimate the strength of residual telluric A band contamination. The red and blue continua are defined to be 7510--7590~\AA and 7670--7750~\AA, respectively, while the ``feature'' band is 7590--7670\AA. The 80\AA~bandwidth is also roughly the bandwidth used for EW(H$\delta$) at $z \sim 0.9$. 
We measure the flux in these three passbands for 221 stars in our spectroscopic catalog, then compare the average flux of red and blue continua to the flux in the ``feature`` bandpass. We find that on average, the flux of the ``feature'' bandpass is $5\%$ ($8\%$) lower than the continuum derived from two adjacent bands for DEIMOS (LRIS) spectra, and only 6\% (14) of the spectra show a deficit of $> 15\%$. 

The underestimated flux in the blue continuum of a galaxy with $z \gtrsim 0.92$ results in an over-estimate of $D_n$(4000). On the other hand, for a galaxy with $z \lesssim 0.91$, its $D_n$(4000) would be underestimated. The worst case scenario is that the entire telluric A band falls in either the red or blue continuum. At $z \sim 0.9$, the 100\AA~bandwidth used for $D_n$(4000) corresponds to 190\AA~in observed frame, which is $> 2$ times broader than the width of the telluric A band. Therefore we estimate that the flux in the affected bandpass is underestimated by $\sim 2\%$ ($4\%$) for DEIMOS (LRIS) spectra, on average. As a result, the $D_n$(4000) will be off by 2\%(4\%) in this scenario. For the $D_n$(4000) measured in the composite spectra, we expect the effect is even lower, because the composite spectra use the full redshift range of Cl1604, $0.84 < z < 0.96$, so the effect on galaxies at higher and lower redshifts roughly cancel each other. Therefore, the effect from the uncorrected telluric A band 
is negligible.

In the case of H$\delta$, only galaxies with $z \lesssim 0.90$ can be affected by the telluric A band. Following the above reasoning, we expect the effect on EW(H$\delta$) of composite spectra would be $\lesssim 5\%$  and thus not a concern. In addition to the composite spectra, we use the EW(H$\delta$) of individual galaxies to select our K+A sample. At the border of our selection criteria, EW(H$\delta$) $>$ 3\AA~, $5\%$ corresponds to 0.15\AA, which is much smaller than our typical measurement error and is therefore negligible.


\begin{thebibliography}{}
\bibitem[Abadi et al.(1999)]{aba99} Abadi, M. G., Moore, B., Bower, R. G. 1999, \mnras, 308, 947
\bibitem[Arnouts et al.(1999)]{arn99} Arnouts, S., Cristiani, S., Moscardini, L., et al. 1999, \mnras, 310, 540
\bibitem[Ascaso et al.(2013)]{asc13} Ascaso, B., Lemaux, B. C., Lubin, L. M., et al. 2013, arXiv 1309.6643
\bibitem[Baldry et al.(2004)]{bal04} Baldry, I. K., Glazebrook, K., Brinkmann, J., et al. 2004, \apj, 600, 681
\bibitem[Baldwin et al.(1981)]{bal81} Baldwin, J. A., Phillips, M. M., Terlevich, R. 1981, \pasp, 93, 5
\bibitem[Balogh et al.(2000)]{bal00} Balogh, M. L., Navarro, J. F., Morris, S. L. 2000, \apj, 540, 113
\bibitem[Balogh et al.(1999)]{bal99} Balogh, M. L., Morris, S. L., Yee, H. K. C., Carlberg, R. G., \& Ellingson, E. 1999, \apj, 527, 54
\bibitem[Barnes \& Hernquist(1992)]{bar92} Barnes, J. E., \& Hernquist., L. 1992, \araa, 30, 705
\bibitem[Bekki et al.(2002)]{bek02} Bekki, K., Couch, W. J., \& Shioya, Y. 2002, \apj, 577, 651
\bibitem[Bekki et al.(2005)]{bek05} Bekki, K., Couch, W. J., Shioya, Y., \& Vazdekis, A. 2005, \mnras, 359, 949
\bibitem[Bell et al.(2004)]{bel04} Bell, E. F., Wolf, C., Meisenheimer, K., et al. 2004, \apj, 608, 752
\bibitem[Bertin et al.(2002)]{ber02} Bertin, E., Mellier, Y., Radovich, M., et al. 2002, ASPC, 281, 228
\bibitem[Biviano et al.(2006)]{biv06} Biviano, A., Murante, G., Borgani, S., et al. 2006, \aap, 456, 23
\bibitem[Blake et al.(2004)]{bla04} Blake, C. et al. 2004, \mnras, 355, 713
\bibitem[Bohlin et al.(1983)]{boh83} Bohlin, R. C., Jenkins, E. B., Spitzer, L., Jr. et al. 1983, \apjs, 51, 277
\bibitem[Bruzual \& Charlot(2003)]{bc03} Bruzual, G., \& Charlot, S. 2003, MNRAS, 344, 1000
\bibitem[Calzetti et al.(2000)]{cal00} Calzetti, D., Armus, L., Bohlin, R. c., et al. 2000, \apj, 533, 682
\bibitem[Casali et al.(2007)]{cas07} Casali M., Adamson, A., Alves de Oliveira, C., et al. 2007, \aap, 467, 777 
\bibitem[Chary \& Elbaz(2001)]{cha01} Chary, R. \& Elbaz, D. 2001, \apj, 556, 562
\bibitem[Choi et al.(2009)]{cho09} Choi, Y., Goto, T., and Yoon, S.-J. 2009, \mnras, 395, 637
\bibitem[Croton et al.(2006)]{cro06} Croton, D. J., Springel, V., White, S. D. M., et al. 2006, \mnras, 365, 11
\bibitem[Dale \& Helou(2002)]{dal02} Dale, D. A. \& Helou, G. 2002, \apj, 576, 159
\bibitem[Davis et al.(2003)]{dav03} Davis, M., Faber, S. M., Newman, J., et al. 2003, SPIE, 4843, 161
\bibitem[Dekel \& Birnboim(2003)]{dek03} Dekel, A., \& Birnboim, Y. 2003, \mnras, 368, 2
\bibitem[Dressler \& Gunn(1983)]{dre83} Dressler, A., \& Gunn, J. E. 1983, \apj, 270, 7 
\bibitem[Dressler et al.(2013)]{dre13} Dressler, A., Oemler, A. Jr., Poggianti, B., et al. 2013, \apj, 770, 62
\bibitem[Dressler et al.(1999)]{dre99} Dressler, A., Smail, I., Poggianti, B. M., et al. 1999, ApJS, 122, 51
\bibitem[Elbaz et al.(2010)]{elb10} Elbaz, D., Hwang, H. S., Magnelli, B., et al. 2010, \aap, 518, L29
\bibitem[Faber et al.(2003)]{fab03} Faber, S. M., Phillips, A. C., Robert, K., et al. 2003, \procspie, 4841, 1657
\bibitem[Faber et al.(2007)]{fab07} Faber, S. M., Willmer, C. N. A., Wolf, C., et al. 2007, \apj, 665, 265
\bibitem[Fisher et al.(1998)]{fis98} Fisher, D., Fabricant, D., Franx, M., \& van Dokkum, P. 1998, \apj, 498, 195
\bibitem[Franzetti et al.(2007)]{fra07} Franzetti, P., Scodeggio, M., Garilli, B. et al, 2007, \aap, 465, 711
\bibitem[Gal \& Lubin(2004)]{gal04} Gal, R. R., Lubin, L. M., 2004, \apj, 607, L1
\bibitem[Gal et al.(2008)]{gal08} Gal, R. R., Lemaux, B. C., Lubin, L. M., Kocevski, D. D., \& Squires, G. K. 2008, \apj, 684, 933
\bibitem[Gal et al.(2005)]{gal05} Gal, R. R., Lubin, L. M., \& Squires, G. K. 2005, \aap, 129, 1827
\bibitem[Goto(2005)]{got05} Goto, T. 2005, \mnras, 357, 937
\bibitem[Goto(2007a)]{got07a} Goto, T. 2007, \mnras, 377, 1222
\bibitem[Goto(2007b)]{got07b} Goto, T. 2007, \mnras, 381, 187
\bibitem[Gunn \& Gott(1972)]{gun72} Gunn, J. E., \& Gott, J. R. I. 1972, \apj, 176, 1
\bibitem[Ilbert et al.(2006)]{ilb06} Ilbert, O., Arnouts, S., McCracken, H. J., et al. 2006, \aap, 457, 841
\bibitem[Juneau et al.(2011)]{jun11} Juneau, S., Dickinson, M., David, M., Salim, S. 2011, \apj, 736, 104
\bibitem[Kaviraj et al.(2007)]{kav07} Kaviraj, S., Kirkby, L., Silk, J., \& Sarzi, M. 2007, \mnras, 382, 960
\bibitem[Kells et al.(1998)]{kel98} Kells, W., Dressler, A., Sivaramakrishnan, A., et al. 1998, \pasp, 110, 1487
\bibitem[Kennicutt(1998)]{ken98} Kennicutt, R. C., Jr. 1998, \araa, 36, 189
\bibitem[Kewley et al.(2004)]{kew04} Kewley, L. J., Geller, M. J., \& Jansen, R. A. 2004, \aap, 127, 2002
\bibitem[Kocevski et al.(2011a)]{koc11a} Kocevski, D. D., Lemaux, B. C., Lubin, L. M., et al. 2011, \apj, 736, 38
\bibitem[Kocevski et al.(2011b)]{koc11b} Kocevski, D. D., Lemaux, B. C., Lubin, L. M., et al. 2011, \apjl, 737, 38
\bibitem[Kocevski et al.(2009)]{koc09} Kocevski, D. D., Lubin, L. M., Gal, R. R., et al. 2009, \apj, 690, 295
\bibitem[Larson et al.(1980)]{lar80} Larson, R. B., Tinsley, B. M., \& Caldwell C. N., 1980, \apj, 237, 692
\bibitem[Le Borgne et al.(2006)]{leb06} Le Borgne, D., Abraham, R., Daniel, K., et al. 2006, \apj, 642, 48
\bibitem[Le F\'{e}vre et al.(2005)]{lef05} Le F\'{e}vre, O., Vettolani, G., Barilli, B. et al. 2005, \aap, 439, 845
\bibitem[Lemaux et al.(2012)]{lem12} Lemaux, B. C., Lubin, L. M., Kocevski, d., et al. 2012, \apj, 745, 106
\bibitem[Lemaux et al.(2009)]{lem09} Lemaux, B. C., Lubin, L. M., Sawicki, M., et al. 2009, \apj, 700, 20
\bibitem[Lemaux et al.(2010)]{lem10} Lemaux, B. C., Lubin, L. M., Shapley, A., et al. 2010, \apj, 716, 970
\bibitem[Lilly et al.(2007)]{lil07} Lilly, S. J., Le F\'evre, O., Renzini, A., et al. 2007, \apjs, 172, 70
\bibitem[Lotz et al.(2008)]{lot08} Lotz, J. M., Jonsson, P., Cox, T. J., \& Primack, J. R. 2008, \mnras, 391, 1137
\bibitem[Lotz et al.(2010a)]{lot10a} Lotz, J. M., Jonsson, P., Cox, T., J., \& Primack, J. R. 2010, \mnras, 404, 575
\bibitem[Lotz et al.(2010b)]{lot10b} Lotz, J. M., Jonsson, P., Cox, T., J., \& Primack, J. R. 2010, \mnras, 404, 590
\bibitem[Lubin et al.(2000)]{lub00} Lubin, L. M., Brunner, R., Metzger, M. R., Postman, M., \& Oke, J. B. 200, \apj, 531, L5
\bibitem[Lubin et al.(2009)]{lub09} Lubin, L. M., Gal, R. R., Lemaux, B. C., Kocevski, D. D., \& Squires, G. K. 2009, \apj, 137, 4867
\bibitem[Ma et al.(2008)]{ma08} Ma, C.-J., Ebeling, H., Donovan, D., \& Barrett, E. 2008, \apj, 684, 160
\bibitem[MacArthur(2005)]{mac05} MacArthur, L. A. 2005, \apj, 623, 795
\bibitem[Mihos \& Hernquist(1994)]{mih94} Mihos, J. C., Hernquist, L. 1994, \apjl, 431, 9
\bibitem[Mihos(1995)]{mih95} Mihos, J. C. 1995, \apjl, 438, 75
\bibitem[Muzzin et al.(2012)]{muz12} Muzzin, A., Wilson, G., Yee, H. K. C., et al. 2012, \apj, 746, 188
\bibitem[Norton et al.(2001)]{nor01} Norton, S. A., Gebhardt, K., Zabludoff, A. I., \& Zaritsky, D. 2001, \apj, 557, 150
\bibitem[Oemler et al.(2013)]{oem13} Oemler, A., Dressler, A., Gladders, M. G., et al, 2013, \apj, 770, 61 
\bibitem[Oemler et al.(2009)]{oem09} Oemler, A., Dressler, A., Kelson, D., et al. 2009, \apj, 693, 152
\bibitem[Oke et al.(1995)]{oke95} Oke, J. B., Cohen, J. G., Carr, M., et al. 1995, \pasp, 107, 375
\bibitem[Peng et al.(2010)]{pen10} Peng, Y.-J., Lilly, S. J., Kova\v{c}, K., et al. 2010, \apj, 721, 193
\bibitem[Poggianti et al.(2009)]{pog09} Poggianti, B. M., Arag\'on-Salamanca, A., Zaritsky, D., et al. 2009, \apj, 693, 112
\bibitem[Poggianti et al.(1999)]{pog99} Poggianti, B. M., Smail, I., Dressler, A., et al. 1999, \apj, 518, 576
\bibitem[Poggianti et al.(2006)]{pog06} Poggianti, B. M., von der Linden, A., de Lucia, G., et al. 2006, \apj, 642, 188
\bibitem[Poggianti \& Wu(2000)]{pog00} Poggianti, B. M., Wu, H. 2000, \apj, 529, 157
\bibitem[Pracy et al.(2005)]{pra05} Pracy, M. B., Couch, W. J., Blake, C., et al. 2005, \mnras, 359, 1421
\bibitem[Pracy et al.(2009)]{pra09} Pracy, M. B., Kuntschner, H., Couch, W. J., et al. 2009, \mnras, 396, 1349
\bibitem[Pracy et al.(2012)]{pra12} Pracy, M. B., Owers, M. S., Couch, W. J., et al. 2012, \mnras, 420, 2232
\bibitem[Quintero et al.(2004)]{qui04} Quintero, A. D., Hogg, D. W., Blanton, M. R., et al. 2004, \apj, 602, 190
\bibitem[Rumbaugh et al.(2012)]{rum12} Rumbaugh, N., Kocevsk, D. D., Gal, R. R., et al. 2012, \apj, 746, 155
\bibitem[Rumbaugh et al.(2013)]{rum13} Rumbaugh, N., Kocevsk, D. D., Gal, R. R., et al. 2013, \apj, 763, 124
\bibitem[Schlegel et al.(1998)]{sch98} Schlegel, D. J., Finkbeiner, D. P., \& Davis, M. 1998, \apj, 500, 525
\bibitem[Simcoe et al.(2000)]{sim00} Simcoe, R. A., Metzger, M. R., Small, T. A., \& Araya, G. 2000, BAAS, 32, 758
\bibitem[Skrutskie et al.(2006)]{skr06} Skrutskie, M. F., Cutri, R. M., Stiening, R., et al. 2006, \aj, 131, 1163
\bibitem[Snyder et al.(2011)]{sny11} Snyder, G. F., Cox, T. J., Hayward, C. C., Hernquist, L., Jonsson, P. 2011, \apj, 741, 77
\bibitem[Springel et al.(2005)]{spi05} Springel, V., Di Matteo, T., \& Hernquist, L. 2005, \mnras, 361, 776
\bibitem[Stasi\'{n}ska et al.(2006)]{sta06} Stasi\'{n}ska, G., Fernandes, R. C., Mateus, A., et al. 2006, \mnras, 371, 972
\bibitem[Strateva et al.(2001)]{str01} Strateva, I., Ivezi\'{c}, \v{Z}., Knapp, G. R., et al. 2001, \apj, 122, 1861
\bibitem[Struck(2006)]{str06} Struck, C. 2006, in Galaxy Collisions-Dawn of a New Era, ed. J. W. Mason (Heidelberg: Springer-Verlag), 115
\bibitem[Surace et al.(2005a)]{sur05a} Surace, J. A., Shupe, D. L., Fang. F., et al. 2005, BAAS, 37, 1246
\bibitem[Surace et al.(2005b)]{sur05b} Surace, J. A., Shupe, D. L., Fang. F., et al. 2005, ``The SWIRE Data Release 2: Image Atlases and Source Catalogs for ELAIS-N1, ELAIS-N2, XMM-LSS, and the Lockman Hole.``, (Pasadena, CA: SSC), \url{http://irsa.ipac.caltech.edu/data/SPITZER/SWIRE/docs/delivery_doc_r2_v2.pdf}
\bibitem[Tody(1986)]{tod86} Tody, D. 1986, Proc. SPIE, 627, 733
\bibitem[Treu et al.(2003)]{tre03} Treu, T., Ellis, R. S., Knieb, J.-P., et al. 2003, \apj, 591, 53
\bibitem[Tran et al.(2003)]{tra03} Tran, L.-V. H., Franx, M., Illingworth, G. D., Kelson, D. D., \& van Dokkum, P. 2003, \apj, 599, 865
\bibitem[Tran et al.(2004)]{tra04} Tran, K.-V. H., Franx, M., Illingworth, G. D., et al. 2004, \apj, 609, 683
\bibitem[Vergani et al.(2010)]{ver10} Vergani, D., Zamorani, G., Lilly, S., et al. 2010, \aap, 509, 42
\bibitem[Wild et al.(2009)]{wil09} Wild, V., Walcher, C. J., Johansson, P. H., et al. 2009, \mnras, 395, 144
\bibitem[Willmer et al.(2006)]{wil06} Willmer, C. N. A., Faber, S. M., Koo, D. C., et al. 2006, \apj, 647, 853
\bibitem[Yan et al.(2006)]{yan06} Yan, R., Newman, J. A., Faber, S. M., et al. 2006, \apj, 648, 281
\bibitem[Yan et al.(2009)]{yan09} Yan, R., Newman, J. A., Faber, S. M., et al. 2009, \mnras, 398, 735
\bibitem[Yang et al.(2004)]{yan04} Yang, Y., Zabludoff, A. I., Zaritsky, D., Lauer, T. R., Mihos, J. C., 2004, \apj, 607, 258 
\bibitem[York et al.(2000)]{yor00} York, D., G., Adelman, J., Anderson, J. E., et al. 2000, \aj, 120, 1579
\end{thebibliography}
\end{document}